\documentclass[%
 aip,amsmath,amssymb,
 reprint,%
]{revtex4-2}

\usepackage{graphicx}
\usepackage{dcolumn}
\usepackage{bm}

\usepackage[mathlines]{lineno}
\usepackage{nicefrac}
\usepackage[textsize=tiny]{todonotes}
\usepackage{cases}
\usepackage{float}
\usepackage{soul}
\setstcolor{red}

\usepackage[utf8]{inputenc}
\usepackage[T1]{fontenc}
\usepackage{mathptmx}
\usepackage{etoolbox}
\usepackage{hyperref}
\renewcommand{\eqref}[1]{Eq.~(\ref{#1})}
\newcommand{\figref}[1]{Fig.~\ref{#1}}
\newcommand{\secref}[1]{Sec.~\ref{#1}}
\newcommand{\appref}[1]{Appendix~\ref{#1}}

\newcommand{\kIB}{k_{\text{I}\rightarrow\text{B}}}
\newcommand{\kBI}{k_{\text{B}\rightarrow\text{I}}}
\newcommand{\zB}{z_\text{B}}
\newcommand{\zI}{z_\text{I}}
\newcommand{\zBT}{\tilde{z}_\text{B}}
\newcommand{\zIT}{\tilde{z}_\text{I}}
\newcommand{\rhov}{\rho_\text{v}}
\newcommand{\rhoB}{\rho_\text{B}}
\newcommand{\rhoI}{\rho_\text{I}}
\newcommand{\rhoET}{\tilde{\rho}_\text{E}}
\newcommand{\rhoBT}{\tilde{\rho}_\text{B}}
\newcommand{\rhoIT}{\tilde{\rho}_\text{I}}
\newcommand{\Dmu}{\Delta\mu}
\newcommand{\Dmucoex}{\Delta\mu_\text{coex}}

\newcommand{\DPhi}{\Delta\Phi}
\newcommand{\Df}{\Delta f}
\newcommand{\Dfl}{\Delta f_\text{l}}
\newcommand{\Dfv}{\Delta f_\text{v}}
\newcommand{\DfT}{\Delta\tilde{f}}

\newcommand{\Dfres}{\Delta f_\text{res}}
\newcommand{\DDf}{\Delta\Delta f}

\newcommand{\Dh}{\Delta h}
\newcommand{\DuIB}{\Delta u_{\text{I}\rightarrow\text{B}}}

\makeatletter
\def\@email#1#2{%
 \endgroup
 \patchcmd{\titleblock@produce}
  {\frontmatter@RRAPformat}
  {\frontmatter@RRAPformat{\produce@RRAP{*#1\href{mailto:#2}{#2}}}\frontmatter@RRAPformat}
  {}{}
}%
\makeatother
\begin{document}

\title{Nonequilibrium interfacial properties of chemically driven fluids}
\author{Yongick Cho}
\author{William M. Jacobs}%
\email{wjacobs@princeton.edu}
\affiliation{Department of Chemistry, Princeton University, Princeton, NJ 08544, USA}

\date{\today}

\begin{abstract}
  Chemically driven fluids can demix to form condensed droplets that exhibit phase behaviors not observed at equilibrium.
  In particular, nonequilibrium interfacial properties can emerge when the chemical reactions are driven differentially between the interior and exterior of the phase-separated droplets.
  Here, we use a minimal model to study changes in the interfacial tension between coexisting phases away from equilibrium.
  Simulations of both droplet nucleation and interface roughness indicate that the nonequilibrium interfacial tension can either be increased or decreased relative to its equilibrium value, depending on whether the driven chemical reactions are accelerated or decelerated within the droplets.
  Finally, we show that these observations can be understood using a predictive theory based on an effective thermodynamic equilibrium.
\end{abstract}

\maketitle

\section{Introduction}

A ``chemically driven'' fluid can maintain a nonequilibrium steady-state (NESS) when provided with a constant external supply of a chemical fuel~\cite{weber2019physics,zwicker2022intertwined}.
By coupling this fuel source to chemical reactions taking place in the fluid, and thereby biasing the forward/backward reaction kinetics, it is possible to drive the fluid away from equilibrium.
A NESS is established when the constant supply of chemical fuel prevents the fluid from relaxing to thermal equilibrium.
For example, in living cells, biomolecules can undergo reactions involving post-translational modifications, such as changes in conformational state due to phosphorylation~\cite{monahan2017phosphorylation,kim2019phosphorylation,nosella2021phosphorylation}.
If the free energy used to drive these transitions away from equilibrium is derived from a constant supply of ATP, then the reaction kinetics will break detailed balance, and the intracellular fluid can be considered to be at a NESS.

Like equilibrium fluids, chemically driven fluids can demix to form coexisting phases via liquid--liquid phase separation~\cite{berry2018physical,weber2019physics}.
However, the phase behavior and phase-separation dynamics of a chemically driven fluid may differ qualitatively from that of a fluid at equilibrium.
For example, coarsening can be suppressed~\cite{zwicker2015suppression}, leading to a monodisperse size distribution of coexisting droplets at steady state~\cite{wurtz2018chemical,kirschbaum2021controlling}, or accelerated~\cite{Tena-Solsona2021ripeningaccelerated} relative to equilibrium.
Nonequilibrium droplets can exhibit self-regulatory behaviors such as self-division~\cite{zwicker2017growth}, self-propulsion~\cite{Demarchi2023Selfpropulsion}, and self-organization into complex internal microstructures that are controlled by the chemical driving forces~\cite{bauermann2022chemical,Donau2022multiphasic}.
The kinetics of droplet assembly and disassembly can also differ from equilibrium phase-separation kinetics and can be controlled by tuning the chemical driving forces~\cite{cho2023nucleation,cates2023nucleation,Ziethen2023nucleation}.
This ability to control the dynamics of droplet formation suggests that nonequilibrium phase separation may be a facile means of achieving spatiotemporal regulation within living cells~\cite{Nott2015PhaseTransition,soding2020mechanisms}.

Understanding how the interfacial tension between coexisting phases changes at a NESS is essential for describing the phase behavior and dynamics of chemically driven fluids~\cite{cho2023nucleation,besse2023interface}.
In particular, \textit{nonequilibrium interfacial tensions}, which imply deviations of the steady-state interfacial properties from equilibrium, emerge when the driven chemical-reaction kinetics differ between the interior and exterior of a phase-separated droplet~\cite{cho2023nucleation}.
Such \textit{spatially inhomogeneous} reaction kinetics can arise in complex fluids when the chemical fuel source is uniformly distributed but the catalyst required for the driven reaction pathway tends to partition into one of the two coexisting phases.
For example, spatially inhomogeneous reaction kinetics can be induced by the preferential enrichment of enzymes either inside or outside of biomolecular condensates in living cells~\cite{bauermann2022chemical,zwicker2022intertwined}.
Since enzymatic partitioning is observed in a wide variety of intracellular condensates as a means of maintaining chemically specific environments to carry out specialized biochemical functions~\cite{oflynn2021role,Schuster2021biocondensate_review}, it is likely that inhomogeneous reaction kinetics---and thus nonequilibrium interfacial properties---are a common feature of nonequilibrium phase-separated droplets in living cells.

In this article, we use a minimal model of a chemically driven fluid to quantify the effects of driven chemical reactions on the interfacial properties between coexisting phases at a NESS.
We show that simulations of droplet nucleation and interface roughness lead to compatible inferences of the nonequilibrium interfacial tension in systems with spatially inhomogeneous chemical reactions.
The rest of this paper is structured as follows:
In \secref{section:model}, we describe the minimal model and consider two chemical-reaction schemes in which the driven reaction is either accelerated or decelerated inside of the droplets.
In \secref{section:thermodynamics}, we discuss how a theory based on an ``effective equilibrium'' can be used to predict both bulk and interfacial properties of the coexisting phases.
Then in \secref{section:interfacial_tension}, we show that the nonequilibrium interfacial tensions inferred from nucleation and roughness simulations agree with the predictions of our effective-equilibrium theory.
Finally, in \secref{section:discussion}, we discuss the implications of our results for conducting noninvasive measurements of interfacial properties in nonequilibrium fluids.

\section{A minimal model for chemically driven fluids}
\label{section:model}

\begin{figure*}
    \centering
    \includegraphics[width=\textwidth]{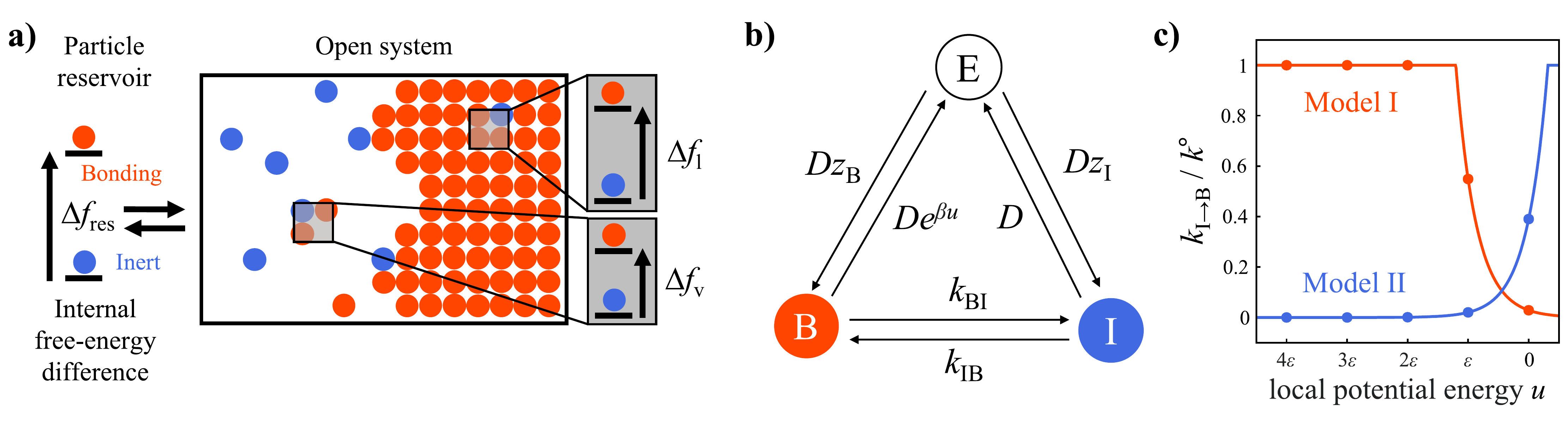}
    \caption{\textbf{Simulating driven chemical reactions at a phase-separated NESS.}
    (a)~Schematic of an open system showing vapor (left) and liquid (right) phases in direct coexistence.  The equilibrium free-energy difference between the bonding (B) and inert (I) states in the particle reservoir is $\Delta f_{\text{res}}$.  The \textit{effective} free-energy differences between the B and I states in the liquid and vapor phases are $\Dfl$ and $\Dfv$, respectively.
    (b)~Kinetic scheme of particle exchange and internal chemical reactions. In simulations of an open system, each lattice site can stochastically transition between being empty (E) or being occupied by either a bonding-state (B) or inert-state (I) particle according to the specified transition rates.
    (c)~Examples of inhomogeneous chemical-reaction models with different dependencies on the local potential energy, $u$, experienced by B-state particles. Orange and blue colors indicate chemical-reaction models I and II, defined by Eqs.~(\ref{eq:decreasing}) and (\ref{eq:increasing}), respectively.
    }
    \label{fig:fig1}
\end{figure*}

In this section, we present a minimal model of a chemically driven fluid, in which reactions allow particles to interconvert between two internal conformational states.
In \secref{subsection:an_open_system}, we describe the modeling approach first introduced in Ref.~\cite{cho2023nucleation}.
We then consider two alternative inhomogeneous chemical-reaction schemes, which result in different reaction kinetics in the condensed and dilute phases, in \secref{subsection:inhomogeneous_chemical_reaction}.
Implementation details of our simulations are presented in \secref{subsection:implementation_of_the_model}.

\subsection{Modeling driven chemical reactions in an open system}
\label{subsection:an_open_system}

In order to investigate interfacial properties in chemically driven systems, we employ a lattice model of a fluid consisting of particles in an implicit solvent~\cite{cho2023nucleation}.
Extending the classical two-dimensional lattice-gas model, we allow particles to undergo chemical reactions between two internal conformational states: a bonding (B) state and an inert (I) state.
Each lattice site can be occupied by at most one particle in either internal state.
Short-ranged attractive interactions between B-state particles tend to drive phase separation, whereas both I-state particles and empty (E) lattice sites, which represent the implicit solvent, are non-interacting.
Specifically, a particle in the B state engages in nearest-neighbor attractive interactions with other neighboring B-state particles with pairwise interaction strength $\epsilon\;(< 0)$.
These interactions give rise to a local potential energy $u = \epsilon b\; (\leq 0)$ at each lattice site, where $b$ is the number of nearest-neighbor lattice sites occupied by B-state particles.
I-state particles, on the other hand, do not interact with nearest-neighbor particles, and are thus isoenergetic to empty lattice sites, with $u = 0$.

Chemical reactions between the B and I conformational states of individual particles occur via two distinct reaction pathways.
An \textit{undriven} pathway is governed by equilibrium thermodynamics, while the \textit{driven} pathway is considered to be coupled to a constant chemical fuel source.
We model these pathways within the framework of stochastic thermodynamics~\cite{seifert2012stochastic,van2015ensemble}, which treats reactions as Markovian events taking place at a constant absolute temperature $T$.
Along the undriven pathway, the ratio of the forward and backward reaction rates is dictated by detailed balance and the internal free-energy difference between the B and I states in an ideal-gas reservoir, $\Dfres$ (\figref{fig:fig1}a).
Throughout this work, we consider scenarios in which $\Dfres > 0$, such that the I state is more stable in the dilute vapor (v) phase, while attractive interactions are required to stabilize the B state in the condensed liquid (l) phase.
The fugacities $\zB$ and $\zI$ of B and I-state particles in an equilibrium ideal gas are thus related to this equilibrium free-energy difference by $\beta\Dfres \equiv -\ln(\zB/\zI)$, where $\beta \equiv (k_\text{B}T)^{-1}$.
Along the driven pathway, conformational-state changes are driven out of equilibrium by a chemical-potential difference $\Dmu$ that originates from the constant chemical fuel source.
The ratio of the forward and backward reaction rates on the driven pathway is thus increased relative to the equilibrium ratio by a factor $\exp(\beta\Dmu)$.
The relative flux between the driven and undriven pathways then determines the steady-state distribution between the conformational states.
In the limit that the flux through the driven pathway vanishes, we recover the equilibrium distribution.
By contrast, if the flux through the driven pathway dominates, then the system approaches an effective equilibrium distribution governed by an internal free-energy difference of $\Dfres + \Dmu$.

We model the fluid as an open system connected to a particle reservoir, which allows us to simulate competition between the driven and undriven pathways under conditions where the reactions are rate-limiting (\figref{fig:fig1}a).
This limit describes a scenario in which particle transport within the fluid is much faster than the typical rate of transitions between conformational states along either pathway.
Importantly, this limit is appropriate for studying interfacial properties that arise purely from the competition between the driven and undriven pathways in a phase-separated system, which is the focus of this paper.
Open systems also provide advantages for studying phase coexistence that are analogous to those of the equilibrium grand-canonical ensemble, such as the elimination of interfaces and reduced finite size effects~\cite{wilding1995critical}.

Within the framework of an open system, we implement reactions along the driven pathway as direct transitions between the B and I states, while reactions along the undriven pathway occur indirectly via particle exchanges with the equilibrium particle reservoir (\figref{fig:fig1}b).
Particle insertion and removal rates depend on the reservoir fugacities, $\zB$ and $\zI$, the local potential energy, $u$, experienced by a B-state particle, and the base exchange rate, $D$, between the system and the reservoir.
Direct transitions between the B and I states, corresponding to driven $\text{I}\rightarrow\text{B}$ and $\text{B}\rightarrow\text{I}$ reactions, proceed with reaction rates $k_\text{IB}$ and $k_\text{BI}$, respectively.
We use dimensionless reaction rates $\kIB \equiv D^{-1}k_\text{IB}$ and $\kBI \equiv D^{-1}k_\text{BI}$ for notational simplicity throughout the rest of the paper.
When the chemical drive $\Dmu$ along the driven pathway is nonzero, the probability of observing a sequence of events around the single-cycle network in \figref{fig:fig1}b differs from that of the time-reversed sequence, breaking detailed balance and leading to a nonzero net probability current.
The chemical drive $\Dmu$ is directly related to the ratio of the probability of observing a forward sequence of transitions around the cycle in the B-to-I direction relative to that of its time-reversed sequence~\cite{seifert2012stochastic, van2015ensemble},
\begin{equation} \label{eq:Dmu}
    \beta\Dmu \equiv \ln\left[\zB\kBI/\zI\kIB e^{\beta u}\right].
\end{equation}
Rearranging \eqref{eq:Dmu} leads to the so-called ``local detailed balance'' condition for the direct-transition rates~\cite{seifert2012stochastic, van2015ensemble},
\begin{equation} \label{eq:local-detailed-balance}
  \kBI/\kIB = \exp(\beta\Dfres + \beta u + \beta\Dmu).
\end{equation}
Thus, \eqref{eq:local-detailed-balance} implies that the ratio of the forward and reversed reaction rates on the driven pathway is increased relative to the ratio on the undriven pathway by a factor $\exp(\beta\Dmu)$, and detailed balance is recovered only when $\Dmu = 0$.

\subsection{Models of inhomogeneous driven chemical reactions}
\label{subsection:inhomogeneous_chemical_reaction}

In this paper, we focus on \textit{inhomogeneous} driven chemical reactions, whose kinetics differ between the liquid and vapor phases.
When the reaction kinetics are identical, or \textit{homogeneous}, in both phases, then the steady-state density distribution of the driven system can be mapped exactly to that of an equilibrium system~\cite{kirschbaum2021controlling}.
Consequently, no change in the interfacial properties is observed when the reaction kinetics are homogeneous~\cite{cho2023nucleation}.
By contrast, when the kinetics differ between the coexisting phases, there is no single effective-equilibrium model that can describe both phases simultaneously.
Coexisting phases in inhomogeneously driven fluids can only be approximately described using \textit{different} effective-equilibrium models, giving rise to nonequilibrium interfacial properties~\cite{cho2023nucleation}.

Inhomogeneous reaction kinetics are implemented in our model by controlling the relative fluxes between the driven and undriven pathways in the two phases.
To this end, we tune the ratio of the fluxes by making the base rate of the driven, direct transitions between the B and I states to be dependent on a particle's local environment within the open system.
Specifically, we set the backward rate $\kIB$ to be dependent on the local potential energy, $u$.
The forward rate, $\kBI$, then follows from the local detailed balance condition, \eqref{eq:local-detailed-balance}.
Here, we consider two chemical reaction schemes in which $\kIB$ takes a Metropolis-like form (\figref{fig:fig1}c):
\begin{numcases}{\kIB=}
    k^\circ \min[1,\exp(- \beta\Dfres - \beta\Dmu - \beta u)] \label{eq:decreasing}\\
    k^\circ \min[1,\exp(-\beta\Dfres - \beta\Dmu + \beta u)] \label{eq:increasing}.
\end{numcases}
The prefactor $k^\circ$ sets the relative flux of the driven pathway compared to the undriven pathway in a dilute vapor phase, where $u \approx 0$.
We refer to the inhomogeneous reaction kinetics defined by Eqs.~(\ref{eq:decreasing}) and (\ref{eq:increasing}) as chemical-reaction models I and II, respectively, throughout the rest of the paper.
In chemical-reaction model I, $\kIB$ decreases monotonically with respect to $u$, implying that the driven reaction is faster---all else being equal---in the liquid phase, where the potential energy tends to be low, than in the vapor phase, where the potential energy tends to be high.
The behavior is the opposite in model II.

\subsection{Implementation via kinetic Monte Carlo algorithm}
\label{subsection:implementation_of_the_model}

To study the consequences of inhomogeneous driven chemical reactions, we perform kinetic Monte Carlo simulations~\cite{gillespie2007stochastic} of our model on a two-dimensional square lattice with periodic boundary conditions (\figref{fig:fig1}a).
We treat particle exchanges and direct transitions between B and I states as first-order Markovian reactions.
We then simulate the stochastic evolution of the lattice with the parameters $\zB$, $\zI$, and $\beta\Dmu$ held constant in time.

All simulation data presented in this paper are obtained at a dimensionless interaction strength of $\beta\epsilon = -2.95$, which is stronger than the critical dimensionless interaction strength, $\beta\epsilon_\text{c} = {-2\ln(1+\sqrt{2})}$,~\cite{pathria1996statistical} of the associated equilibrium lattice-gas model.
Two of the three control parameters, $\zB$, $\zI$, and $\beta\Dmu$, must then be specified in order to define a unique coexistence point.
We choose parameters for simulation by controlling both the extent of the nonequilibrium drive, $\beta\Dmu$, and the average number density of particles in either internal state in the vapor phase, $\rhov$.
The latter of these conditions is achieved by tuning the internal free-energy difference $\Dfres$; focusing on $\rhov$ as opposed to $\Dfres$ makes for an easier connection to experimental settings.
Throughout this work, we consider coexistence points at which the vapor-phase number density is $\rhov = 0.05$.
Finally, the prefactor for the driven chemical-reaction pathway, $k^\circ$, is set to $k^\circ = 0.1$ for chemical-reaction model I and $k^\circ = 1$ for model II.
These choices yield relative timescales for the driven and undriven pathways in which the interfacial properties clearly deviate from equilibrium, as opposed to the behavior in the undriven-reaction-dominated ($k^\circ\rightarrow0$) or driven-reaction-dominated ($k^\circ\rightarrow\infty$) limits~\cite{cho2023nucleation}.

\section{Chemical reaction-induced nonequilibrium phase behavior at coexistence}
\label{section:thermodynamics}

\begin{figure*}
    \centering
    \includegraphics[width=\textwidth]{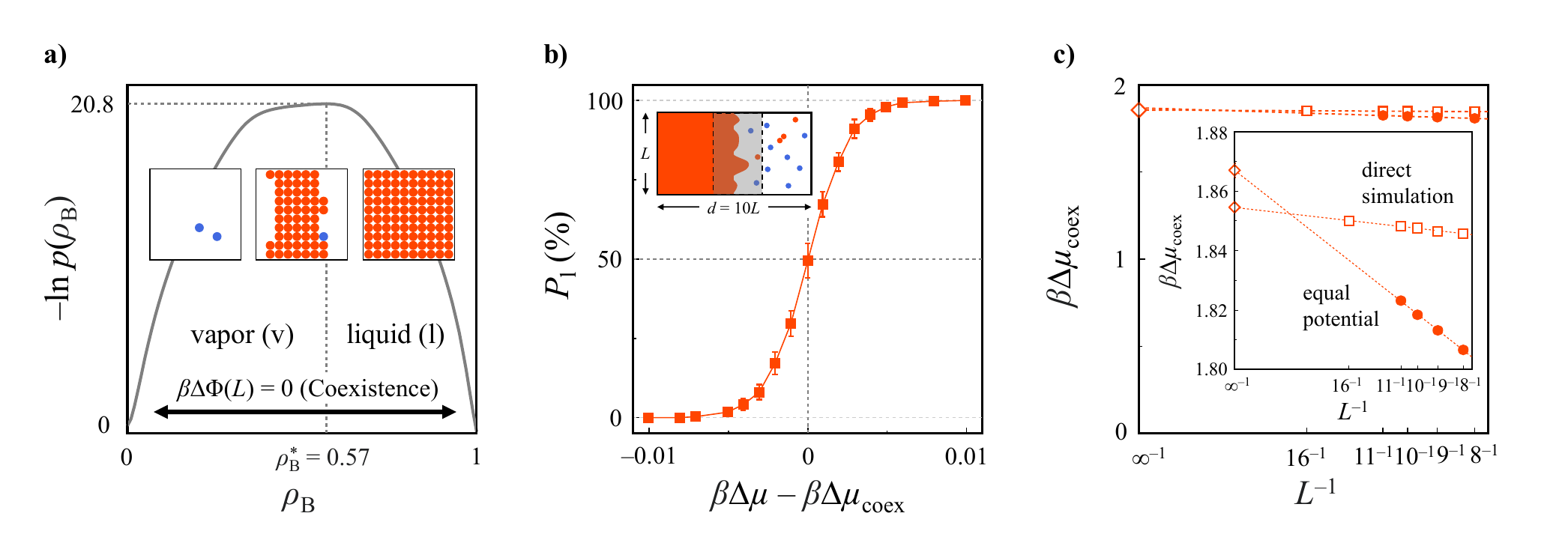}
    \caption{\textbf{Definitions of nonequilibrium phase coexistence.}
      (a)~To define phase coexistence based on the steady-state density distribution, $p(\rhoB)$, we equate the nonequilibrium potential, $\Phi$, in the liquid and vapor phases (see text).  The liquid and vapor phases are defined by the location of the top of the barrier, $\rhoB^*$, with respect to $-\ln p(\rhoB)$. In this example, the steady-state density distribution is shown for lattice dimension $L = 10$.
      \textit{Insets:} Representative lattice configurations are shown, from left to right, in the vapor phase; at the top of the barrier, $\rho = \rhoB^*$; and in the liquid phase. Colors indicate particle internal states as in \figref{fig:fig1}a.
      (b)~To define phase coexistence based on direct-coexistence simulations, we measure the probability, $P_{\text{l}}$, that the liquid phase expands to fill the entire lattice starting from a planar liquid--vapor interface.  Coexistence occurs when $P_\text{l}$ equals 50\%. In this example, calculations are shown for lattice dimension $L = 16$.
      \textit{Inset:} Schematic of the $L \times d$ setup for direct-coexistence simulations. Simulations are performed within the $L \times L$ shaded window, which tracks the center of the planar liquid--vapor interface.
    (c)~The two definitions of phase coexistence from panels (a) and (b) match closely in the macroscopic limit, $L \rightarrow \infty$.  Filled circles and open squares indicate coexistence points determined from the definitions in panels (a) and (b), respectively.  Open diamonds indicate the extrapolated coexistence points in the macroscopic limit.  All data are obtained using chemical-reaction model I.}
    \label{fig:fig2}
\end{figure*}

In this section, we demonstrate the effect of driven chemical reactions on the bulk phase behavior of the coexisting phases, which sets the stage for examining nonequilibrium interfacial properties.
In \secref{subsection:direct_coexistence}, we establish the condition for phase coexistence under nonequilibrium conditions based on mechanical balance, providing further validation of the approach presented in Ref.~\cite{cho2023nucleation}.
We then examine how the nonequilibrium phase behavior depends on the chemical-reaction models defined in \secref{subsection:thermodynamics}, and we show how these results can be understood in terms of a theory based on an effective-equilibrium approximation, which was originally introduced in Ref.~\cite{cho2023nucleation}, in \secref{subsection:FLEX}.

\subsection{Determination of phase coexistence}
\label{subsection:direct_coexistence}

Two phases are in coexistence when there is no net energy or mass flux between them and they are mechanically balanced~\cite{chandler1987introduction}.
At equilibrium, a variational principle based on the second law of thermodynamics implies that equal temperatures, chemical potentials, and pressures between the coexisting phases lead to energy, mass, and mechanical balance, respectively, and vice versa.
In the grand-canonical ensemble at equilibrium, in which the temperature and chemical potential are fixed, ensuring that the grand-potential densities are the same in the two phases satisfies the equal-pressure condition~\cite{hansen2013simpleliquids}.
Although our model is driven out-of-equilibrium, the particle reservoir still guarantees that the energy and mass conditions are satisfied by providing the energy and mass input required to keep each phase at steady-state.
However, equating the grand-potential densities in the two phases does not ensure mechanical balance in fluids that are out of equilibrium due to the general absence of an equation of state~\cite{solon2015pressure,brady2023mechanical}.

We therefore examine two closely related definitions of nonequilibrium phase coexistence based on mechanical balance, which are not guaranteed to coincide in the case of chemically driven fluids \textit{a priori}.
First, we determine phase coexistence by equating the steady-state probability of observing the open system in either the liquid or the vapor phase.
We then compare this definition of coexistence with the result of nonequilibrium direct-coexistence simulations.

In the steady-state distribution approach, we calculate the steady-state number-density distribution of the B-state particle, $p(\rhoB)$, on the domain $\rhoB \in [0,1]$.
This calculation is performed using an $L\times L$ lattice.
To this end, we employ a form of nonequilibrium umbrella sampling (NEUS)~\cite{warmflash2007umbrella} with respect to a one-dimensional collective variable, $\rhoB$, that we have developed previously~\cite{cho2023nucleation}.
In this method, we divide the one-dimensional $\rhoB$ axis into non-overlapping windows, calculate the steady-state distribution within each window, and enforce detailed balance between the adjacent windows in order to reconstruct the entire steady-state distribution.
We observe that the system exists with high probability at either low or high values of $\rhoB$, corresponding to the vapor and liquid phases, respectively (\figref{fig:fig2}a).
The phases are separated by a high barrier with respect to $-\ln p(\rhoB)$, as expected for a first-order phase transition~\cite{chandler1987introduction}.
The height of this barrier scales approximately linearly with the lattice dimension, $L$, since the barrier reflects the formation of a liquid--vapor interface on the two-dimensional lattice.
Based on the location of the top of the barrier, $\rhoB^*$, we determine the probability of being in the vapor phase, $p_\text{v} \equiv \int_0^{\rhoB^*} p(\rhoB)d\rhoB$, versus the liquid phase, $p_\text{l} \equiv \int_{\rhoB^*}^1 p(\rhoB)d\rhoB$.
Due to the distinct nature of the phases and the extremely small steady-state probability near the top of the barrier, the precise definitions of $p_\text{v}$ and $p_\text{l}$ are insensitive to the choice of collective variable and the location of $\rhoB^*$.
We then define the nonequilibrium potential difference, $\DPhi$, per lattice site between the liquid and vapor phases, $\beta\DPhi(L) \equiv L^{-2}\ln(p_\text{l}/p_\text{v})$.
The potential difference in the macroscopic limit, $\DPhi_\infty \equiv \lim_{L\rightarrow\infty} \DPhi(L)$, is exactly equal to the difference between the grand-potential densities of the two phases in a fluid at equilibrium~\cite{Wilding1992critical}.
We therefore associate $\DPhi_\infty = 0$ with nonequilibrium phase coexistence in this approach.
This operational definition of $\DPhi$ is particularly useful, as it plays the role of a thermodynamic potential for the liquid--vapor phase transition under nonequilibrium conditions and can be evaluated directly from NEUS simulations.

In the direct-coexistence approach~\cite{noya2008}, we consider an open system that is periodic in the vertical direction but always in contact with the liquid and vapor phases in the horizontal direction (inset of \figref{fig:fig2}b).
Fluctuations at the liquid--vapor interface cause the portion of the lattice that is occupied by the liquid phase to expand or shrink.
Starting from an initial configuration in which the lattice is split equally between the liquid and vapor phases, we simulate the system until the liquid phase either fills the lattice entirely or vanishes completely.
For simulation efficiency, we only allow particle exchanges and chemical reactions to take place within a window of size $L\times L$, which tracks the position of the planar liquid--vapor interface (inset of \figref{fig:fig2}b); the width of this window is much larger than the interface height fluctuations, which ensures that the window boundaries do not affect the growth or shrinkage of the liquid phase.
We then measure the probability that the lattice is occupied by the liquid phase at the end of the simulation, $P_\text{l}$.
Consistent with the notion of mechanical balance~\cite{brady2023mechanical}, we associate phase coexistence with the condition $P_\text{l} = 50\%$.

We find that the coexistence points identified by these two approaches match closely when extrapolated to the macroscopic limit.
In both approaches, we find that the chemical drive at phase coexistence, $\Dmucoex$, shows a consistent dependence on the lattice dimension, $L$, which allows us to extrapolate the coexistence condition to the macroscopic limit, $L\rightarrow\infty$ (\figref{fig:fig2}c).
Interestingly, the $1/L$ scaling that we observe is reminiscent of effective Coulombic interactions predicted in other treatments of chemically reactive systems~\cite{Christensen1996coulombic,Ziethen2023nucleation}.
Even in the case of a strongly driven system $(|\beta\Dmu| \gg 1)$, the two definitions of phase coexistence yield very similar coexistence points, $\Dmucoex$ (differing by less than $10^{-2} k_\text{B}T$), in the macroscopic limit (\figref{fig:fig2}c).
These differences are typically orders of magnitude smaller than the chemical drive at coexistence, as well as the range of $\Dmu$ values over which simulate droplet nucleation (see \secref{subsection:nucleation}).
We therefore conclude that these two coexistence definitions agree well within the range of system parameters that we consider here, and we utilize the definition $\beta\DPhi_\infty = 0$ for operational convenience.

\subsection{Effective-equilibrium quantification of nonequilibrium phase behavior}
\label{subsection:thermodynamics}

\begin{figure*}
    \centering
    \includegraphics[width=\textwidth]{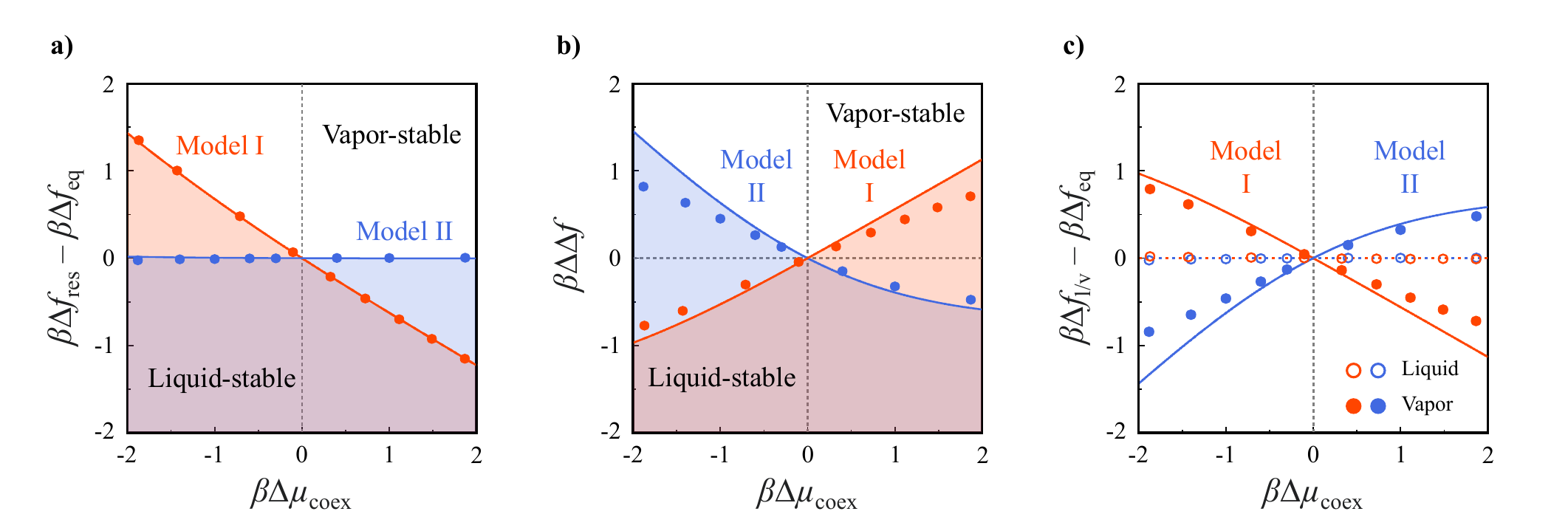}
    \caption{\textbf{Nonequilibrium phase behavior of coexisting bulk phases.}
    (a)~The phase diagram of the open system depends on the inhomogeneous chemical-reaction scheme, as illustrated by the differences between chemical-reaction models I (orange) and II (blue).  In both cases, coexistence is calculated for a vapor phase with a number density of $\rhov = 0.05$.
    (b)~Driven chemical reactions induce different effective internal free-energy differences, $\DDf \equiv \Dfl - \Dfv$, between the coexisting phases (see text).
    (c)~In the inhomogeneous chemical-reaction schemes considered in this work, the effective internal free-energy differences arise primarily due to changes in the vapor phase (solid symbols) as opposed to the liquid phase (open symbols).
    In all panels, symbols indicate results obtained from simulations of bulk phases at the coexistence point, while lines indicate theoretical predictions obtained using FLEX.
    Shaded (unshaded) regions in panels (a) and (b) indicate where the liquid (vapor) phase is predicted to be stable by FLEX.
    Solid and dotted lines in panel (c) indicate the vapor and liquid phases, respectively.
}
    \label{fig:fig3}
\end{figure*}

Because the relative fluxes depend on the kinetics of the chemical reactions, reaction models I and II lead to different coexistence lines, as shown in \figref{fig:fig3}a.
In this phase diagram, $\Dfres$ decreases with $\Dmucoex$ along the coexistence line in the case of model I, while the value of $\Dfres$ at coexistence is essentially independent of the chemical drive in the case of model II.
In both cases, the total number density in the vapor phase is fixed such that $\rhov = 0.05$ at coexistence.

The differences between the phase diagrams of the two chemical-reaction models suggest that the bulk properties of the coexisting phases depend on the details of the reaction models at a given nonequilibrium chemical drive.
To understand how the driven chemical reactions affect the bulk properties of the coexisting liquid and vapor phases, we map each phase to an \textit{effective-equilibrium} model.
Here, effective equilibrium implies an equilibrium fluid with the same B-state interaction strength, $\epsilon$, as in the nonequilibrium system, but with a potentially different \textit{effective} internal free-energy difference between the B and I states, $\Delta f$.
This effective free-energy difference is calculated using a generalized Widom insertion formula~\cite{Allen2017liquid,cho2023nucleation},
\begin{equation} \label{eq:eff_eq_df}
  \beta\Df \equiv -\ln(\rhoB/\rhoI) + \ln\langle\exp(-\beta\DuIB)\rangle_\text{I},
\end{equation}
where $\rhoB$ and $\rhoI$ are the steady-state number densities of B and I-state particles, respectively, in a bulk phase; $\Delta u_{\text{I}\rightarrow\text{B}}$ is the potential energy change due to changing an I-state particle to the B state; and the angle brackets $\langle \cdot \rangle_{\text{I}}$ indicate a steady-state average conditioned on a tagged lattice site being occupied by an I-state particle.
In equilibrium, Eq.~\ref{eq:eff_eq_df} reduces to $\beta\Df_\text{eq} = \beta\Dfres$, where $\Df_\text{eq}$ is the equilibrium internal free-energy difference.
At a nonzero $\Dmucoex$, we evaluate \eqref{eq:eff_eq_df} from a steady-state trajectory in which the lattice is occupied entirely by one phase.
In this way, we are able to determine effective free-energy differences, $\Dfl$ and $\Dfv$, for the coexisting liquid and vapor phases, respectively (\figref{fig:fig1}a).
We note that while this effective equilibrium mapping is not exact in general, it nevertheless offers useful insights and provides a quantitative foundation for understanding and predicting nonequilibrium phase behaviors (see \appref{app:test_eff_eq}).

In general, coexisting phases cannot be described by the same effective-equilibrium model in driven fluids with inhomogeneous chemical reactions.
It is therefore useful to quantify the \textit{difference} between the effective internal free-energy differences in the coexisting liquid and vapor phases, $\DDf \equiv \Dfl - \Dfv$.
In \figref{fig:fig3}b, we observe that as the system is driven farther away from equilibrium along a coexistence line, $\DDf$ either increases or decreases monotonically, signaling growing differences between the effective thermodynamics of the coexisting phases as the magnitude of $\beta\Dmucoex$ is increased.
However, the relationship between $\beta\DDf$ and $\beta\Dmucoex$ depends on the details of the inhomogeneous chemical-reaction model.
We find that $\beta\DDf$ increases with $\beta\Dmucoex$ in the case of model I, while the relationship is reversed in the case of model II.
The physical origin of this behavior is discussed in the following section.

Unexpectedly, we find that the deviations of $\beta\DDf$ from zero are almost entirely due to the vapor phase for both chemical-reaction models that we examine here.
\figref{fig:fig3}c shows that $\beta\Dfv$ either decreases or increases monotonically in the case of models I or II, respectively.
By contrast, $\Dfl$ exhibits an almost negligible change at the coexistence conditions.
A theoretical explanation for these observations is discussed below.

\subsection{FLEX description of nonequilibrium phase behavior}
\label{subsection:FLEX}

To understand the effect of driven chemical reactions on the nonequilibrium phase behavior, we utilize the ``Fixed Local Environment approXimation" (FLEX) theoretical framework introduced in Ref.~\cite{cho2023nucleation} (see \appref{app:FLEX}).
The central assumption of FLEX is that fluctuations of the local potential energy, $u$, experienced by a B-state particle are small at the level of an individual lattice site.
This assumption is well justified in a bulk phase (see \appref{app:test_eff_eq}) and turns out to provide qualitatively accurate predictions of interfacial properties as well.
Assuming that we can neglect such fluctuations in the local ``environment'' around a lattice site allows us to map the steady-state at a tagged lattice site to an effective equilibrium at constant $u$.
In this way, FLEX predicts that the effective internal free-energy difference, $\Df(u)$, of a lattice site in an environment where a B-state particle would experience a local potential energy $u$ is given by (see \appref{app:FLEX})
\begin{equation}
\label{eq:df_FLEX}
\beta\Df(u) = \beta\Delta{f}_\text{res} + \ln\left[\dfrac{1+\kIB(u)(1+e^{\beta\Delta{f}_\text{res}})e^{\beta\Dmu}}{1+\kIB(u)(1+e^{\beta\Delta{f}_\text{res}})}\right].
\end{equation}
This equation implies that the effective internal free-energy difference between the B and I states at a tagged lattice site is the sum of the equilibrium internal free-energy difference in the particle reservoir and the influence of the driven chemical-reaction pathway.

In order to employ FLEX without resorting to simulation, we assume that $u \approx 4\epsilon$ in the liquid phase, meaning that every lattice is surrounded by B-state particles, and $u \approx 0$ in the vapor phase, meaning that particles are sparsely distributed (see \appref{app:test_eff_eq}).
\eqref{eq:df_FLEX} then predicts that $\DDf = \Dfl - \Dfv = \Df(4\epsilon) - \Df(0)$ is an increasing function of the chemical drive when $\kIB(4\epsilon) > \kIB(0)$, as in model I, while the relationship is the opposite when $\kIB(4\epsilon) < \kIB(0)$, as in model II.
These qualitative predictions agree with the observed simulation results presented in \figref{fig:fig3}b.
Thus, we attribute the qualitative differences in $\beta\DDf$ at coexistence to the acceleration of the driven chemical reactions in either the liquid or vapor phase in the case of chemical-reaction models I or II, respectively (\figref{fig:fig1}c).
Quantitative predictions of the bulk-phase properties obtained from FLEX also agree well with the simulation results at phase coexistence (solid lines in \figref{fig:fig3}; see \appref{app:FLEX-models}).

\section{Effects of driven chemical reactions on nonequilibrium interfacial properties}
\label{section:interfacial_tension}

In this section, we demonstrate that driven chemical reactions can induce changes in the interfacial properties of a nonequilibrium phase-separated fluid relative to those of an equilibrium fluid.
We first infer the nonequilibrium interfacial tension from simulations of droplet-nucleation kinetics in \secref{subsection:nucleation}, extending the results first presented in Ref.~\cite{cho2023nucleation}.
To further validate the conclusions of Ref.~\cite{cho2023nucleation}, we now perform these calculations along two independent supersaturation pathways to establish that the interfacial tension is a material property determined by the coexistence point.
We then show how changes in the interfacial tension that arise in both chemical-reaction models can be understood within the FLEX framework in \secref{subsection:interfacial_tension_change}.
Finally, in \secref{subsection:roughness}, we demonstrate that the nonequilibrium interfacial tension inferred from nucleation simulations is consistent with the height fluctuations of the liquid--vapor interface at phase coexistence.
The results of this section provide a new, independent test of the interfacial-tension measurements reported in Ref.~\cite{cho2023nucleation}.

\subsection{Inferring a nonequilibrium interfacial tension using classical nucleation theory}
\label{subsection:nucleation}

\begin{figure}
    \centering
    \includegraphics[width=0.5\textwidth]{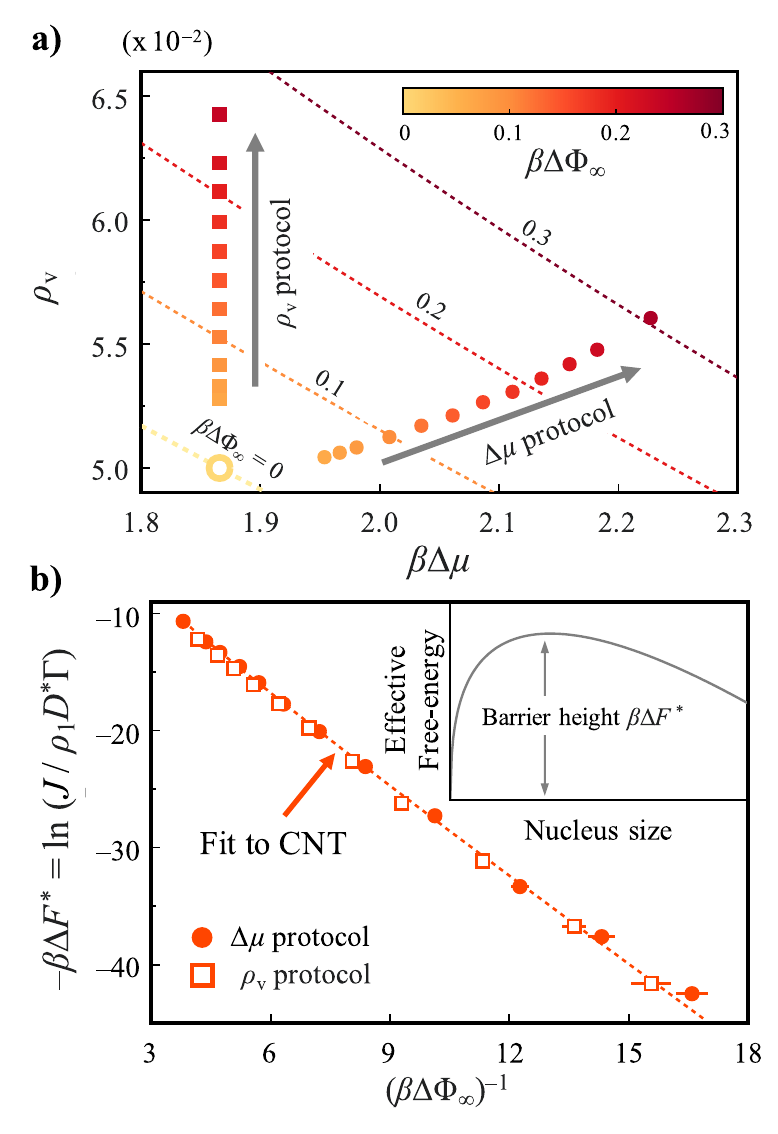}
    \caption{\textbf{Different supersaturation protocols yield compatible inferences of the nonequilibrium interfacial tension in simulations of droplet nucleation.}
    (a)~When supersaturating the vapor phase beyond the coexistence point, the nonequilibrium thermodynamic driving force, $\DPhi_\infty$, can be controlled either by tuning $\Dmu$ or $\rhov$.  Shown here are two such protocols, which vary either $\rhov$ (squares) or $\beta\Dmu$ (circles).  The coexistence point at $(\beta\Dmucoex, \rhov) = (1.87,0.05)$ is indicated by an empty circle, and dotted lines show FLEX predictions of equi-$\DPhi_\infty$ contours.  In this example, the chemical-reaction kinetics follow model I.
    (b)~The interfacial tension inferred from the nucleation kinetics is essentially independent of the protocol used to supersaturate the vapor phase.  The effective free-energy barrier, $\beta\Delta F^*$ (see inset and refer to the text for details), is determined at multiple values of the supersaturation, $\beta\Delta\Phi_\infty$.  Symbols correspond to the protocols shown in panel (a).  The dotted line is a fit of the observed nucleation kinetics to classical nucleation theory (CNT), from which the nonequilibrium interfacial tension, $\sigma_{\text{CNT}}$, is inferred.}
    \label{fig:fig4}
\end{figure}

We first infer the nonequilibrium interfacial tension using simulations of droplet nucleation (\figref{fig:fig4}a), which we interpret using classical nucleation theory (CNT)~\cite{cho2023nucleation}.
When applied to equilibrium fluids, CNT predicts that the nucleation of a liquid droplet from a supersaturated, metastable vapor phase follows the minimum free-energy pathway along a reaction coordinate corresponding to the number of particles, $n$, in a cluster.
This cluster represents a ``nucleus'' of the stable liquid phase (inset of \figref{fig:fig4}b).
The height of the nucleation barrier on this free-energy landscape, $\Delta F^*$, is determined by a competition between the thermodynamic driving force favoring the phase transition to the liquid phase and the free-energy cost of forming the liquid--vapor interface for the liquid droplet.
At a NESS, this thermodynamic driving force must be replaced by an appropriate measure of the supersaturation with respect to the nonequilibrium coexistence point, as we discuss below.
If $\Delta F^*$ is much larger than typical thermal fluctuations, such that $\beta\Delta F^* \gg 1$, then the nucleation rate is dominated by the rate of crossing this barrier.
Under such conditions, CNT predicts that the nucleation rate density, $J$, follows the phenomenological law~\cite{oxtoby1992homogeneous}
\begin{equation}
  J = \rho_1 D^* \Gamma \exp(-\beta\Delta F^*),
\end{equation}
where the height of the free-energy barrier enters in the form of a Boltzmann factor.
Three terms comprise the kinetic prefactor: the number density of B-state monomers that  are not engaged in any attractive interactions, $\rho_1$; the one-dimensional diffusion coefficient along the reaction coordinate near the top of the nucleation barrier, $D^*$; and the Zeldovich factor, $\Gamma$, which accounts for barrier re-crossing events due to fluctuations along the reaction coordinate before a nucleation attempt ultimately succeeds or fails.
CNT has been shown to provide a quantitative description of nucleation kinetics in two-dimensional lattice-gas models at equilibrium~\cite{ryu2010validity}.

We compute the nucleation rate density, $J$, starting from a supersaturated vapor phase using forward-flux sampling (FFS)~\cite{allen2009forward}.
While calculating $J$, we also determine the commitment probability for a successful nucleation event, $\phi(n)$, along the reaction coordinate.
We then compute the Zeldovich factor by fitting $\phi(n)$ to a harmonic free-energy landscape in the vicinity of the top of the nucleation barrier where, $n = n^*$~\cite{Makarov2010Harmonic,cho2023nucleation}, at which the commitment probability equals 50\%~\cite{Hummer2004Transitionpath}.
The remaining quantities $\rho_1$ and $D^*$ are independently determined from simulations of the bulk vapor phase and the evolution of the nucleus size along a trajectory initiated from the critical nucleus size, $n^*$, respectively.
From these measurements, we compute $\beta\Delta F^* \equiv -\ln(J / \rho_1 D^* \Gamma)$, which corresponds to the effective free-energy barrier height for a nonequilibrium nucleation process.

We then apply CNT to our model of a fluid with driven chemical reactions by introducing a nonequilibrium interfacial tension, $\sigma_\text{CNT}$.
To this end, we assume that the nucleation behavior also admits an effective-equilibrium description~\cite{cates2023nucleation}.
We therefore fit the observed effective barrier heights, $\beta\Delta F^*$, as a function of the supersaturation, $\exp(\beta\DPhi_\infty)$, to an analytical expression for the two-dimensional lattice-gas free-energy landscape~\cite{ryu2010validity} using $\sigma_\text{CNT}$ as the sole fitting parameter~\cite{cho2023nucleation}.
This approach assumes that the thermodynamic driving force for the nucleation process is given by $\DPhi_\infty$, which is consistent with the fundamental CNT assumption that both the bulk and the interfacial properties of a nucleus are characteristics of the fluid in the macroscopic limit.

We now consider how the CNT description of the nucleation kinetics depends on the manner in which a chemically driven fluid is supersaturated.
In an equilibrium lattice-gas model at constant interaction strength, the thermodynamic driving force can be increased only by increasing the total particle density in the vapor phase, $\rhov$.
In fluids with driven chemical reactions, however, the chemical drive, $\Dmu$, acts as an additional degree of freedom and thus provides an alternative means of tuning the supersaturation.
We therefore consider two different supersaturation protocols (\figref{fig:fig4}a), in which we either increase $\zB$ and $\zI$ under fixed $\Dmu$ and $\Dfres$, or increase $\Dmu$ with $\Dmu+\Dfres$ fixed [see Eqs. (\ref{eq:decreasing}) and (\ref{eq:increasing})].
The former ``$\rhov$ protocol'' increases the particle density in the vapor phase while keeping $\Dmu$ and the relative flux between the reaction pathways constant.
By contrast, the latter ``$\Dmu$ protocol'' changes the nonequilibrium potential by tuning the chemical drive along the driven reaction pathway.
In both supersaturation protocols, we keep the reaction-rate, $\kIB$, fixed.
We also use FLEX to predict equi-$\Delta\Phi_\infty$ contours in the $(\beta\Dmu,\rhov)$ plane (\figref{fig:fig4}a).
Viewing the parameter space in this way shows that an entire family of supersaturation protocols could plausibly be defined by monotonically tuning $\beta\Dmu$ and $\rhov$ away from a coexistence point, where $\beta\DPhi_\infty = 0$.

Remarkably, despite the differences between the $\rhov$ and $\Dmu$ supersaturation protocols, we find that the nucleation kinetics can be described by CNT to near quantitative accuracy using a common value of the nonequilibrium interfacial tension, $\sigma_{\text{CNT}}$.
This result can be seen in \figref{fig:fig4}b, where we plot the effective nucleation barrier as a function of the supersaturation along the two supersaturation protocols shown in \figref{fig:fig4}a.
We emphasize that the fitted value of $\sigma_\text{CNT}$ differs substantially from the equilibrium interfacial tension in the example shown in \figref{fig:fig4}b, consistent with the fact that the fluid is driven away from equilibrium by inhomogeneous chemical reactions.
The good agreement between the barrier measurements along both protocols, as well as the fit to CNT, indicates that the effective nucleation barriers along both protocols are determined by the same value of the nonequilibrium interfacial tension.
We therefore propose that the nonequilibrium interfacial tension is relatively independent of the supersaturation protocol, as long as the degree of supersaturation is low ($\beta\DPhi_\infty \ll 1$) and the protocols converge at the same coexistence point.
This result agrees with a fundamental premise of CNT that $\sigma_\text{CNT}$ is identical to the interfacial tension of a macroscopic interface between two coexisting phases~\cite{oxtoby1992homogeneous}.
Thus, for the fitting parameter $\sigma_{\text{CNT}}$ to be considered to be a nonequilibrium interfacial tension, it is important that its value is determined by the coexistence point alone.
Our simulation results along these two supersaturation protocols suggest that this is indeed the case.

\subsection{Dependence of the nonequilibrium interfacial tension on the chemical drive at phase coexistence}
\label{subsection:interfacial_tension_change}

\begin{figure}
    \centering
    \includegraphics[width=0.5\textwidth]{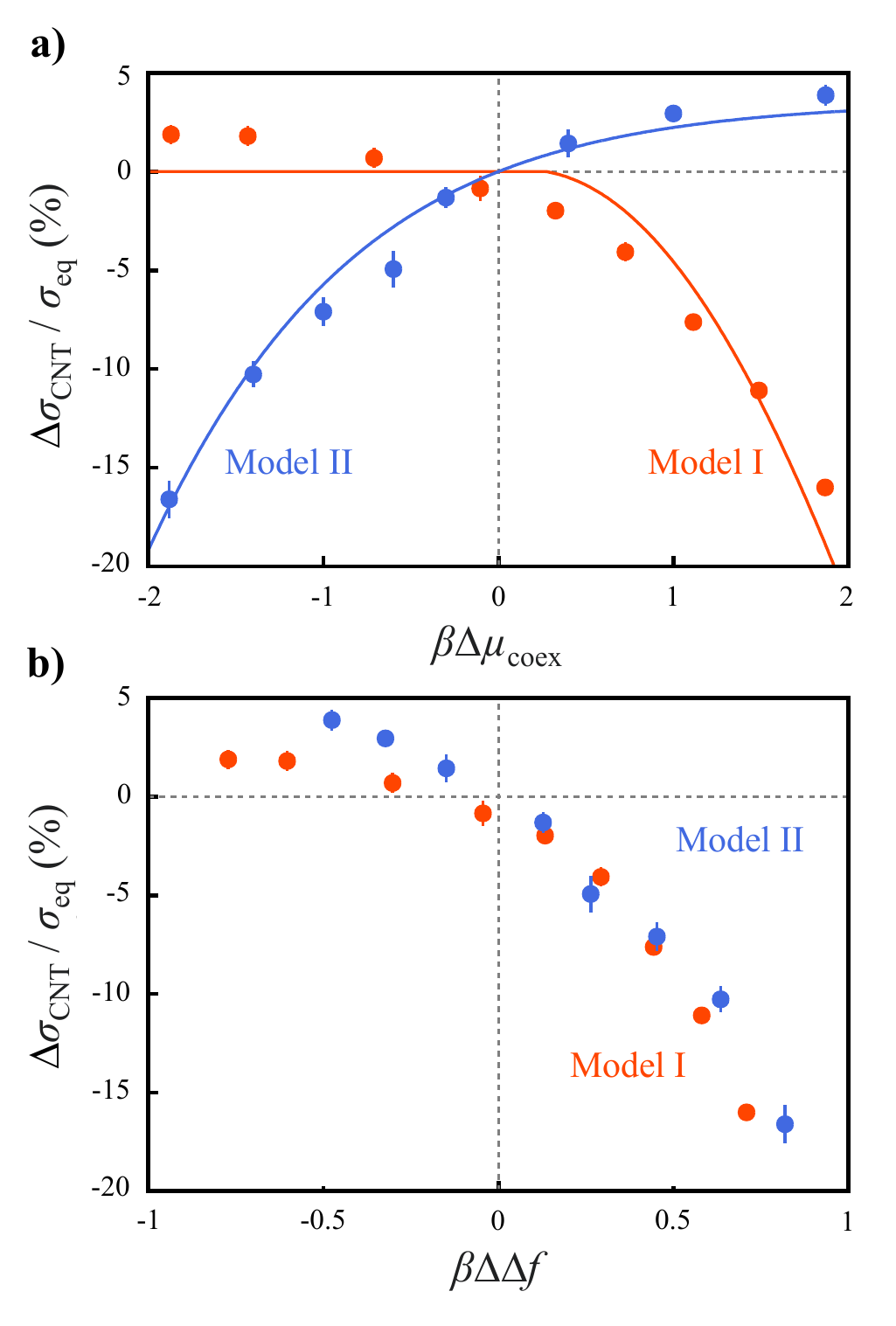}
    \caption{\textbf{Inhomogeneous driven chemical-reaction kinetics determine the nonequilibrium interfacial tension.}
    (a)~The deviation of the nonequilibrium interfacial tension from the equilibrium value, $\Delta\sigma_\text{CNT}$, as a function of the coexistence point at $\rhov = 0.05$.  Results are shown for both inhomogeneous chemical-reaction models I and II.  Solid curves indicate theoretical predictions from FLEX.
    (b)~Increasing the relative stability of B-state particles in the vapor phase ($\beta\DDf > 0$) reduces the nonequilibrium interfacial tension relative to equilibrium, while decreasing the relative stability $(\beta\DDf < 0)$ has the opposite effect.  This relationship holds regardless of the inhomogeneous chemical reaction-kinetics scheme.}
    \label{fig:fig5}
\end{figure}

We find that the nonequilibrium interfacial tension inferred from droplet-nucleation simulations is a function of the chemical drive at coexistence, $\Dmucoex$ (\figref{fig:fig5}a).
As the fluid is driven further away from equilibrium, we find that the interfacial tension determined from nucleation kinetics, $\sigma_\text{CNT}$, tends to deviate further from the equilibrium value, $\sigma_\text{eq}$.
However, the manner in which $\sigma_{\text{CNT}}$ changes as a function of $\Dmucoex$ differs between the two chemical-reaction models.
In the case of model I, $\sigma_{\text{CNT}}$ \textit{increases} monotonically with respect to $\Dmucoex$, while it monotonically \textit{decreases} in the case of model II.
Thus, knowing whether the chemical drive biases the driven reaction pathway in either the $\text{I}\rightarrow\text{B}$ or the $\text{B}\rightarrow\text{I}$ direction is insufficient to predict whether the nonequilibrium interfacial tension will increase or decrease relative to equilibrium.

Instead, the effective-equilibrium parameter $\DDf$ offers a common explanation of the trends in the nonequilibrium interfacial tension for both chemical-reaction models (\figref{fig:fig5}b).
This observation implies that the effects of the driven chemical reactions on the interfacial properties are related to the differences between the effective-equilibrium behaviors of the coexisting phases.
This relationship can be explained by considering the relative populations of B and I-state particles at the interface between the coexisting liquid and vapor phases.
We first recall that a low value of $\Df$ indicates that the B state is more stable than the I state in terms of its internal free-energy; thus, decreasing $\Df$ increases the probability of finding a particle in the B state at a given $u$.
If $\DDf = \Dfl - \Dfv$ is positive, then the effective internal free-energy difference is lower in the vapor phase than in the liquid phase, meaning that the population of B-state particles is higher in the vapor than what would be expected on the basis of an effective-equilibrium description of the liquid phase.
A negative $\DDf$ implies the opposite behavior.
Because $\kIB$ monotonically changes with $u$ for both chemical-reaction models, the sign of $\DDf$ also indicates whether the B-state particle is more or less populated at the liquid--vapor interface than would be expected on the basis of the liquid phase.
Increasing the population of B-state particles at the interface relative to the liquid phase, at a positive $\DDf$, has the effect of lowering the effective free-energy cost of forming the interface.
As a result, a positive $\DDf$ implies a reduced interfacial tension relative to the equilibrium value.
By the same logic, a negative $\DDf$ implies an increase in the interfacial tension relative to the equilibrium value.
These qualitative arguments agree with the observed changes in the interfacial tension presented in \figref{fig:fig5}b.

The effective-equilibrium approach therefore allows us to explain how the interfacial tension changes along the coexistence line for a given chemical-reaction model.
As noted above, $\DDf$ is predominantly determined by the behavior of the vapor phase in both chemical-reaction models studied here (\figref{fig:fig3}c).
Thus, in the case of chemical-reaction model I, $\Dfv$ decreases monotonically with respect to $\Dmucoex$, and the B-state particle population increases, in the vapor phase.
The B-state population at the interface is affected in the same way, so that the interfacial tension decreases with the applied chemical drive.
In the case of model II, on the other hand, $\Dfv$ increases with the applied chemical drive, so that the interfacial tension shows the opposite dependence.
These contrasting trends agree with our simulation findings in \figref{fig:fig5}a.

These arguments are also borne out quantitatively within the FLEX framework (see \appref{app:FLEX-interface}).
FLEX assumes that B-state particles at a flat interface experience a local potential energy of $u = \epsilon$, which lies between $u \approx 4\epsilon$ and $u \approx 0$ for the liquid and vapor phases, respectively.
Because $\kIB$ monotonically changes with $u$ for both chemical-reaction models, \eqref{eq:df_FLEX} predicts that $\Df(\epsilon)$ lies between $\Dfl$ and $\Dfv$, so that a positive (negative) $\DDf$ also indicates an increase (decrease) in the B-state particle population at the interface relative to the liquid phase.
FLEX then predicts that the interfacial tension can be determined by defining an effective dimensionless interaction energy at the interface, $\beta\tilde\epsilon$, which can differ from the actual dimensionless interaction energy, $\beta\epsilon$.
When $\DDf$ is positive (negative), the relative increase (decrease) in the B-state particles at the interface means that the effective interaction energy needed to adsorb B-state particles to the interface is decreased (increased) relative to $\beta\epsilon$.
The nonequilibrium interfacial tension can then be predicted on the basis of the effective interaction energy, $\beta\tilde\epsilon$, at the interface (see \appref{app:FLEX-interface}).
These FLEX predictions are shown as solid lines in \figref{fig:fig5}a.

\subsection{Inferring a nonequilibrium interfacial tension using direct-coexistence simulations}
\label{subsection:roughness}

We further examine the relationship between driven chemical reactions and the nonequilibrium interfacial tension by focusing on the roughness of the liquid--vapor interface.
Since our interpretation of the nucleation kinetics points to nonequilibrium effects at the droplet interface, we hypothesize that changes in the interfacial properties should be observable in simulations of liquid and vapor phases in direct coexistence.
In order to verify this, we test whether changes in the interfacial tension match changes in the height fluctuations of the liquid phase at the interface, which are directly related to the interfacial tension at equilibrium.
If the effective interaction strength increases, then intuition based on equilibrium phase behavior suggests that the interface should become smoother due to the greater effective free-energy cost of introducing curvature at the interface.
Similarly, we anticipate that rougher interfaces will be observed when the interfacial tension is lower.

\begin{figure}
    \centering
    \includegraphics[width=0.5\textwidth]{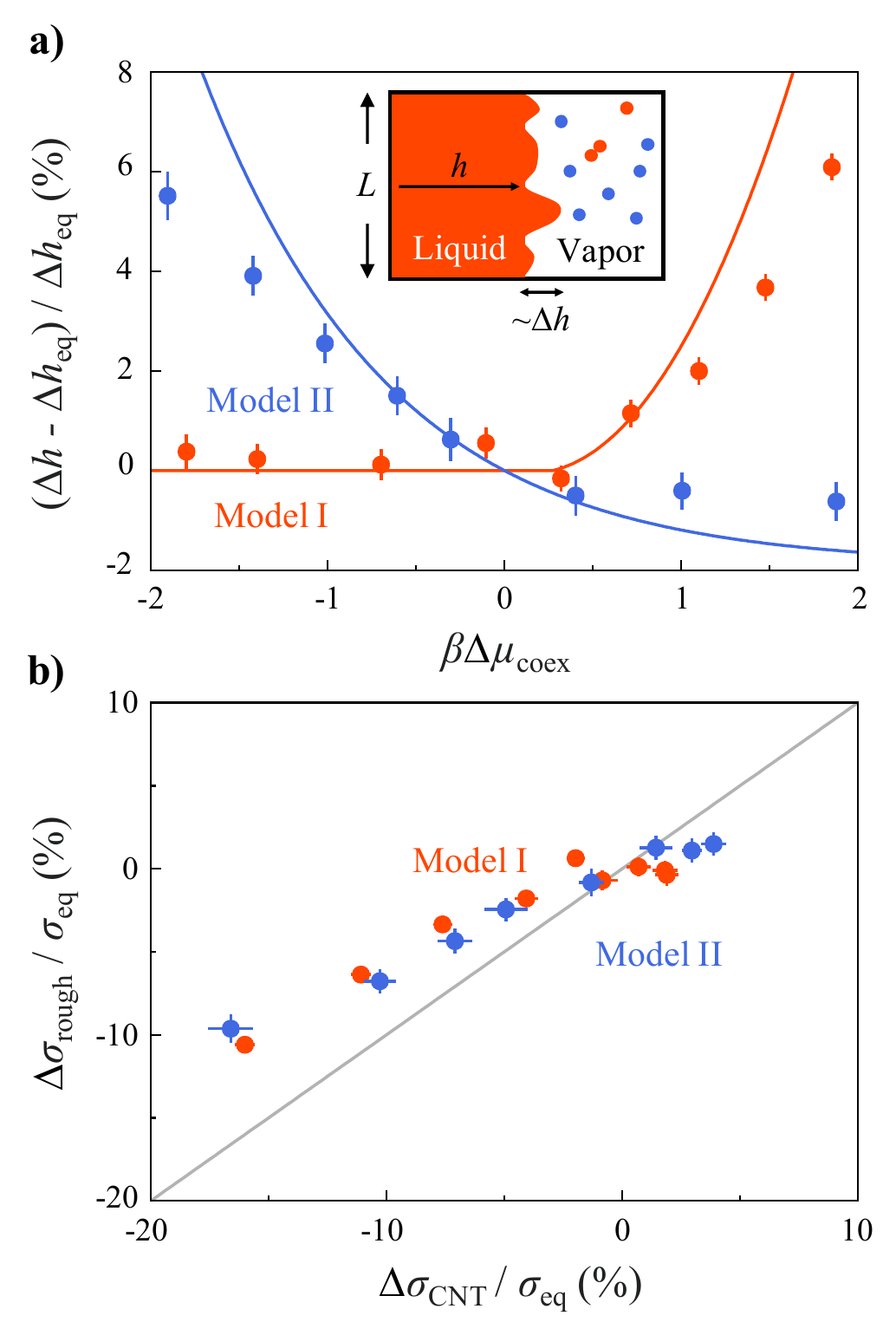}
    \caption{\textbf{Measurements of interfacial roughness at phase coexistence are consistent with the nonequilibrium interfacial tension inferred from nucleation kinetics.}
    (a)~Change in the interfacial roughness, $\Dh$, at coexistence, compared to the equilibrium value, $\Dh_\text{eq}$.
    Symbols show the average of 96 independent simulations of a planar interface with lattice dimension $L = 64$, and error bars indicate standard errors of mean.
    Solid lines are theoretical predictions from FLEX.
    \textit{Inset:} Schematic of the interfacial roughness, $\Dh$, at coexistence.
    (b)~Comparison between the nonequilibrium interfacial tension inferred from the interfacial roughness, $\sigma_\text{rough}$, and from nucleation kinetics, $\sigma_\text{CNT}$.
    Orange and blue colors indicate inhomogeneous chemical-reaction models I and II, respectively.}
    \label{fig:fig6}
\end{figure}

In our direct-coexistence simulations, we focus on the location of the planar liquid--vapor interface in the normal direction.
During the simulation, points on the interface, $\{h\}$, defined as the left-most boundary of all rows that are filled with B-state particles (inset of \figref{fig:fig6}a), fluctuate as B-state particles attach and detach from the liquid phase.
We then define the interfacial roughness, $\Dh$, to be the standard deviation of the heights $\{h\}$,
\begin{equation}
  \Dh \equiv \langle(h - \langle h \rangle)^2\rangle^{1/2},
\end{equation}
where the angle brackets denote an average over all $L$ rows of the lattice.
This definition of $\Dh$ ignores overhanging particles on sides of the particle rows.
We simulate the interface between the coexisting phases on an $L \times L$ lattice that is equally divided between liquid and vapor phases, in the same manner as described in \secref{subsection:direct_coexistence}.
Due to the system-size dependence of the coexistence point (\figref{fig:fig2}c) and the maximum length scale of fluctuations imposed by the finite lattice dimension $L$~\cite{saito1996statistical}, the interfacial roughness is inherently subject to finite-size effects.
We therefore perform simulations at the $L$-dependent coexistence point of the finite-sized lattice with $L = 64$, as opposed to the coexistence point in the macroscopic limit.

We interpret the roughness of the liquid--vapor interface using the equilibrium solid-on-solid (SOS) approximation~\cite{saito1996statistical}.
In this approximation, B-state particles are assumed to attach to the interface only at the top of the particle rows, without overhanging neighboring rows.
Assuming that the interface can be described using the equilibrium SOS approximation, the observed roughness, $\Dh$, can be related to the equilibrium value, $\Dh_\text{eq}$, via
\begin{equation} \label{eq:roughness_ratio}
  \dfrac{\Delta h}{\Delta h_\text{eq}} = \dfrac{\sinh(\beta\epsilon/4)}{\sinh(\beta\tilde{\epsilon}/4)},
\end{equation}
where $\beta\tilde{\epsilon}$ is the effective dimensionless interaction strength at the interface under the SOS approximation.
We can then obtain the effective interfacial tension, $\sigma_\text{rough}(\beta\tilde{\epsilon})$, by solving for $\beta\tilde{\epsilon}$ using \eqref{eq:roughness_ratio} and applying an analytical formula for the equilibrium lattice-gas model [see \eqref{eq:sigma_analytical} in \appref{app:FLEX-interface}].

Analyzing the interfacial roughness within the framework of the SOS model broadly supports our hypothesis of a chemically driven nonequilibrium interfacial tension.
In \figref{fig:fig6}a, we show a comparison between the change in the roughness relative to equilibrium, $(\Delta h - \Delta h_{\text{eq}}) / \Delta h_{\text{eq}}$, and predictions based on FLEX.
For these theoretical predictions, we use the FLEX expression for the effective dimensionless interaction strength, $\beta\tilde\epsilon$ [see \eqref{eq:FLEX_epsilon} in \appref{app:FLEX-interface}], in combination with the SOS relation, \eqref{eq:roughness_ratio}.
The direct-coexistence measurements and FLEX predictions are consistent in that the interfacial roughness increases with $\Dmucoex$ in the case of chemical-reaction model I, while it decreases in the case of model II.
These results are also consistent with the equilibrium intuition that the interfacial roughness increases when the interfacial tension decreases, and vice versa.

To directly compare the results of inference methods based on nucleation kinetics and interfacial roughness, we plot the relationship between $\sigma_\text{CNT}$ and $\sigma_\text{rough}$ in \figref{fig:fig6}b.
Each point in this comparison corresponds to a unique nonequilibrium coexistence point, $\beta\Dmucoex$.
Although the dynamic range of $\sigma_{\text{rough}}$ tends to be smaller than that of $\sigma_{\text{CNT}}$, which is likely a consequence of the SOS approximation used in the former measurement, we observe a strong correlation between these two independent inferences of the nonequilibrium interfacial tension.
Importantly, this correlation holds equally well for both chemical-reaction models.
We therefore conclude that both inference methods yield compatible measurements of the nonequilibrium interfacial tension, and that this interfacial property is entirely determined by the nonequilibrium coexistence conditions.

\section{Discussion}
\label{section:discussion}

In this paper, we show that driven chemical reactions alter the material properties of interfaces between coexisting phases in nonequilibrium fluids.
To this end, we consider two models of inhomogeneous chemical-reaction kinetics, which are either accelerated or decelerated in the condensed liquid phase relative to the dilute vapor phase, using a minimal lattice model~\cite{cho2023nucleation}.
We find that the interfacial tension, determined either from measurements of droplet-nucleation kinetics or from measurements of fluctuations at a planar interface, changes as the system is driven out of equilibrium.
Our interpretation of this effect as a nonequilibrium interfacial tension is supported by the strong correlation between these two independent measurements.
However, the relationship between the nonequilibrium interfacial tension and the applied chemical drive depends on the details of the chemical-reaction model.

We can explain the observed deviations in the interfacial properties using a theoretical framework based on effective-equilibrium models of the coexisting phases.
We interpret our simulation results by mapping each phase in a nonequilibrium fluid to an effective-equilibrium model whose particle density distribution is closest to that of the NESS under a constant chemical drive.
By calculating the effective free-energy differences between the internal states of particles in both phases, we find that we can explain the bulk-phase properties of driven fluids with different chemical-reaction models.
Then, by applying this effective-equilibrium approach to predict the effective free-energy cost of forming a liquid--vapor interface, we obtain a consistent explanation for the observed trends in the nonequilibrium interfacial tensions with different chemical-reaction models.

The concept of effective equilibrium thus plays an essential role in our analysis of the effects of driven chemical reactions.
Nonetheless, we emphasize that our simulation approach, which is based on stochastic thermodynamics, and the underlying lattice model do not make any assumptions regarding an effective equilibrium.
Instead, we utilize the concept of an effective equilibrium only when interpreting our simulation results or when generating approximate analytical predictions using FLEX.

Alternatively, the concept of an effective equilibrium can be employed as a starting point for modeling the thermodynamic and transport properties of chemically driven fluids.
For example, in mesoscale modeling approaches based on linear irreversible thermodynamics~\cite{weber2019physics,zwicker2022intertwined}, the NESS is obtained starting from a free-energy functional that implies local equilibration within a fluid element~\cite{kirschbaum2021controlling,cates2023nucleation,Ziethen2023nucleation}.
In such models, therefore, local effective equilibrium is assumed to provide a complete description of the system, from which stationary states and phase-transformation pathways can be computed, rather than an approximation for analyzing nonequilibrium phenomena as in this paper.
However, this premise may not be justified in nonequilibrium fluids with driven chemical reactions, since the predictions of the effective equilibrium may be incorrect at molecular length scales (see \appref{app:test_eff_eq}).
Models that treat fluid elements at a mesoscopic scale may therefore not capture effects arising as a result of molecular-scale deviations from the effective-equilibrium assumption.
We expect the differences between these approaches will become more substantial as the system approaches a critical point, since fluctuations in the local environment may become too large for fluid elements to be described adequately by a single effective equilibrium.

Although we have studied the effects of driven chemical reactions in a reaction-limited regime in this paper, we expect that our qualitative predictions of nonequilibrium interfacial properties will also hold beyond this regime.
Extending our modeling approach to examine the influence of particle diffusion on nonequilibrium interfacial properties thus represents a promising direction for future study.
Studying nonequilibrium interfacial properties may also be possible within the context of linear irreversible thermodynamics~\cite{besse2023interface}, although local effective-equilibrium assumptions are integral to such models as noted above.
However, we note that any study of interfacial properties in mean-field models of chemically driven fluids~\cite{kirschbaum2021controlling,zwicker2022intertwined} will need to disentangle interfacial effects from those arising from changes in the coexistence conditions.
For example, distinguishing between interfacial and bulk effects is important for comparing nucleation behavior between stochastic thermodynamics-based~\cite{cho2023nucleation} and mean-field approaches~\cite{cates2023nucleation,Ziethen2023nucleation}.

Our simulations suggest that measurements of interfacial roughness can serve as a general method for identifying nonequilibrium effects in driven fluids.
In particular, our analysis indicates that the interfacial tension under different nonequilibrium coexistence conditions can be inferred quantitatively regardless of the kinetics of driven chemical reactions.
This noninvasive approach is particularly appealing, as it could be applied directly to living samples~\cite{Schuster2021biocondensate_review} and to ``aging'' systems whose material properties change over time~\cite{Jawerth2020aging}.
We note that a related strategy has been employed for human cell nuclei fusion processes, in which the interfacial tension is determined from the fluctuation--dissipation theorem~\cite{Caragine2018Surface}.
This approach is similar to our proposal, as both strategies are based on the concept of an effective equilibrium; however, our simulations suggest that our inference approach could potentially be used under far-from-equilibrium conditions, where $|\beta\Dmucoex| > 1$.

Finally, we note that recently developed experimental settings for studying chemically driven fluids could provide suitable platforms for testing our predictions.
To realize the inhomogeneous driven chemical reactions that are essential for observing nonequilibrium interfacial properties, the reaction kinetics along the driven pathway could be tuned by controlling the partitioning of enzymes into phase-separated droplets~\cite{saleh2020enzymatic} or by engineering the reactions to be dependent on the local concentration of droplet material~\cite{spaeth2021molecular,nakashima2021active}.
Investigations of interface roughening could then be used to probe the nonequilibrium phase behavior of these experimental systems.
Since we anticipate that interface roughening could be observed most easily near the critical point, it will also be necessary to inspect the critical behavior of inhomogeneously driven fluids.
Such studies present another important direction for future investigation via theory and simulation.

\begin{acknowledgments}
The authors thank Sushant Saryal for helpful discussions and comments on the manuscript.
This work is supported by the National Science Foundation (DMR-2143670).
\end{acknowledgments}

\section*{Author Declarations}
\subsection*{Conflict of Interest Statement}
The authors have no conflicts to disclose.

\appendix

\section{Testing the effective-equilibrium mapping}
\label{app:test_eff_eq}

\begin{figure*}
    \centering
    \includegraphics[width=\textwidth]{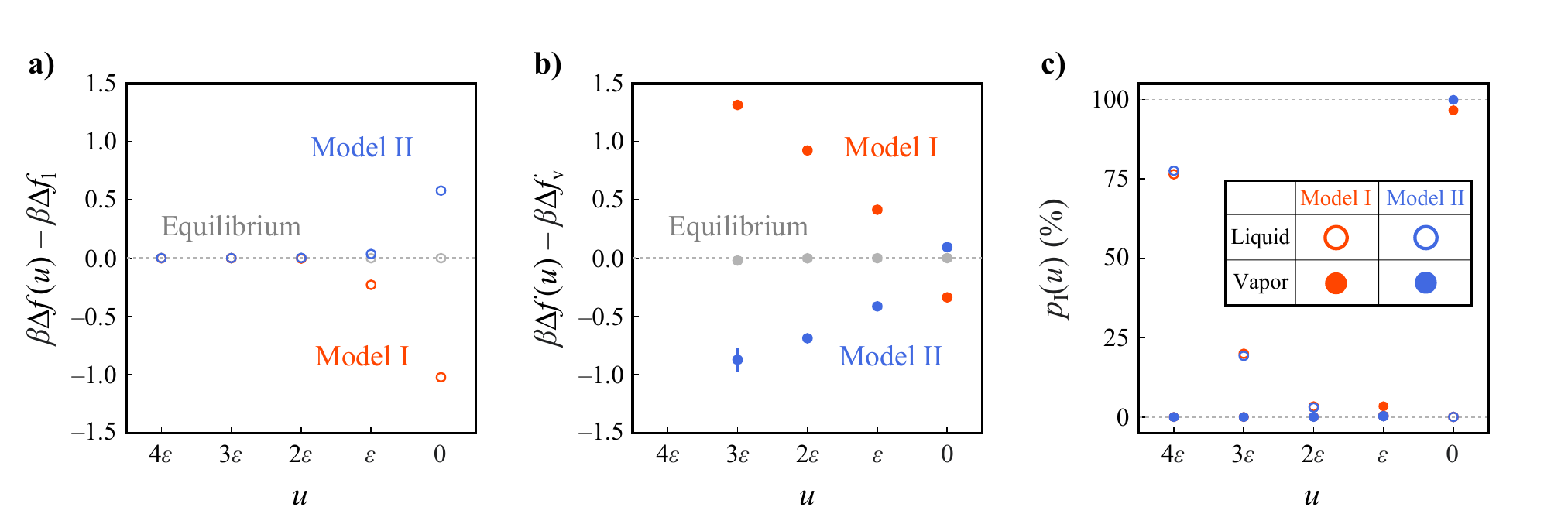} \vskip-1.5ex
    \caption{\textbf{Chemically driven nonequilibrium fluids cannot be mapped exactly to effective-equilibrium models.}
    (a--b)~In fluids with driven chemical reactions, the effective internal free-energy difference defined at the level of an individual lattice site, $\Df(u)$, depends on the local potential energy, $u$, experienced by a B-state particle.  The value of $\Df(u)$ may differ from the bulk-phase values, (a)~$\Dfl$ in the liquid phase and (b)~$\Dfv$ in the vapor phase, at coexistence.  At equilibrium, however, lattice configurations obey the Boltzmann distribution, and the two definitions of the internal free-energy difference coincide.
    (c)~The distribution of the local environment around I-state particles, $p_\text{I}(u)$, in the coexisting liquid (open symbols) and vapor phases (solid symbols).  The local environment is characterized by the local potential energy, $u$, that would be experienced if the particle were in the B state.  Gray, orange, and blue colors indicate equilibrium and driven systems with chemical-reaction models I and II $(\beta\Dmucoex=1.87)$, respectively.
    In (b), results for $u = 4\epsilon$ are not presented, as such configurations are extremely rare in the vapor phase.}
    \label{fig:fig7}
\end{figure*}

In this section, we quantitatively assess the mapping between a nonequilibrium bulk phase and the corresponding effective-equilibrium model.
To this end, we consider an effective internal free-energy difference defined at a microscopic level with the length scale of an individual lattice site, $\Df(u)$, and compare this with the ensemble-averaged effective internal free-energy difference, $\Df$, defined in \eqref{eq:eff_eq_df}.
We define $\Df(u)$ by mapping lattice sites at which a B-state particle experiences a local potential energy $u$ to an effective equilibrium in which the ratio of the number densities for the B and I-states follows the Boltzmann distribution, i.e., $(\rhoB/\rhoI)_u = \exp[-\beta\Df(u)-\beta u]$.
Equivalently, we can write
\begin{equation} \label{eq:eff_eq_df_lattice}
    \beta\Df(u) \equiv -\ln(\rhoB/\rhoI)_u - \beta u,
\end{equation}
where the subscript $u$ emphasizes that the number densities are measured at lattice sites where B-state particles would experience precisely this potential energy.
This definition is closely related to the effective internal free-energy difference for a bulk phase given by \eqref{eq:eff_eq_df}, which may be considered to be an average of \eqref{eq:eff_eq_df_lattice}, weighted by the distribution of $u$ across lattice sites in that phase.

\figref{fig:fig7}a--b show that the internal free-energy differences at the lattice-site level, $\Df(u)$, depend on the local potential energy and are not always equal to the bulk-phase values, $\Dfl$ and $\Dfv$, under nonzero nonequilibrium drive for both chemical-reaction models.
By contrast, the inferred internal free-energy difference is independent of $u$ at equilibrium.
This inconsistency between \eqref{eq:eff_eq_df} and \eqref{eq:eff_eq_df_lattice} demonstrates that the effective-equilibrium mappings for both the liquid and the vapor phases are not exact, suggesting that exact mappings to effective-equilibrium models cannot be defined for nonequilibrium fluids in general.

Despite the inexact nature of the effective-equilibrium approach, the approximate mapping defined by \eqref{eq:eff_eq_df} nonetheless provides a useful description of the macroscopic phase behavior because most particles in a given bulk phase are subject to similar local environments.
\figref{fig:fig7}c shows the distribution of the local environment around I-state particles, $p_\text{I}(u)$, where $u$ indicates the local potential energy that would be experienced if the particle were in the B state.
This distribution is sharply peaked at $u=4\epsilon$ in the liquid phase and at $u=0$ in the vapor phase.
Consequently, we are justified in making the approximations $\Dfl \approx \Df(4\epsilon)$ and $\Dfv \approx \Df(0)$ in our approximate FLEX theory (see \appref{app:FLEX}).
Effective equilibrium is thus a useful conceptual tool as long as the particles in each phase experience similar local environments.

This effective-equilibrium mapping can be successfully applied to chemically driven fluids far from a critical point, where the assumption of a constant local environment in each bulk phase is valid, so that a single effective-equilibrium model provides a reasonable description of each bulk phase.
However, the concept of an effective equilibrium may no longer be as useful for fluids near a critical point or when modeling processes such as spinodal decomposition, where fluctuations cannot be ignored.

\section{Fixed Local Environment approXimation (FLEX)}
\label{app:FLEX}

In the following sections, we reproduce the description of the ``Fixed Local Environment approXimation" (FLEX) derived in our previous work~\cite{cho2023nucleation} for the reader's convenience.
We refer the reader to Ref.~\cite{cho2023nucleation} for further details.
In FLEX, we assume that fluctuations of of the local configuration, or ``environment'', surrounding a tagged lattice site are negligible.
We describe the local environment in terms of the the potential energy, $u$, experienced by a B-state particle at the tagged lattice site.
In keeping with this small-fluctuation assumption, we expect that FLEX will be most effective in systems that are far from a critical point, as discussed in \appref{app:test_eff_eq}.

We use FLEX to map our nonequilibrium model to an effective equilibrium with identical particle number densities $\tilde{\rho}_\text{B}$ and $\tilde{\rho}_\text{I}$.
Given a specified $u$, we calculate the steady-state distribution, $\tilde{\rho}_i$ ($i = $ E, B, or I), within a tagged site site by solving the master equation for the Markovian transition network presented in \figref{fig:fig1}b.
Given the assumption of a fixed local environment, $\rhoBT$ and $\rhoIT$ may regarded as the number densities of particles in B and I states at the steady-state.
The tagged site is then mapped to an effective-equilibrium model with identical B and I-state number densities by defining the effective fugacities $\zBT(u) \equiv (\rhoBT/\rhoET)_u\exp(\beta u)$ and $\zIT(u) \equiv (\rhoIT/\rhoET)_u$, where the single-site partition function is $\tilde{\xi}(u) = 1+\zBT(u)+\zIT(u)$.

Because the number densities depend on $u$, the liquid and vapor phases, which are characterized by different average potential energy, are mapped to distinct effective equilibria.
We quantify the difference in the effective thermodynamic properties of these phases in terms of an effective internal free-energy difference, $\beta\DfT(u) \equiv -\ln(\zBT/\zIT)_u$.
Evaluating this quantity with the effective fugacities leads to \eqref{eq:df_FLEX} in the main text.
The FLEX prediction for the internal free-energy difference captures the general trend of how each phase deviates from thermal equilibrium as the system is driven away from equilibrium.
This equation also highlights the importance of controlling the functional dependence of $\kIB$ with respect to the local potential energy in order to realize inhomogeneous chemical reactions.

FLEX illustrates the roles of $\Dmu$ and $\Dfres$ in tuning the relative flux of the two competing reaction pathways between the particle states.
The flux through the direct $\text{I}\rightarrow\text{B}$ pathway is $j_{\text{I}\rightarrow\text{B}} = \rhoIT\kIB$, while the flux through the indirect $\text{I}\rightarrow\text{E}\rightarrow\text{B}$ pathway is $j_{\text{I}\rightarrow\text{E}\rightarrow\text{B}} = \rhoIT/(1+e^{\beta\Dfres})$.
Thus, the relative flux between the two pathways is
\begin{equation}
\label{eq:FLEX_relative_flux}
j_{\text{I}\rightarrow\text{B}}/{j_{\text{I}\rightarrow\text{E}\rightarrow\text{B}}} = \kIB(1+e^{\beta\Dfres}).
\end{equation}
FLEX predicts that along the coexistence line, specified by $\Dfres$ and $\Dmu$ (\figref{fig:fig3}a), this relative flux gradually changes, leading to deviations in the interfacial properties and phase-transition kinetics relative to an equilibrium fluid.

FLEX makes quantitative predictions of the phase behavior of the open system at a given nonequilibrium drive, $\Dmu$, by assuming that the particle--hole symmetry of two-dimensional lattice-gas model is a reasonable approximation at the effective equilibrium.
In the equilibrium lattice-gas model, the liquid and vapor phases coexist at $\mu = 2\epsilon$, where $\mu$ is the chemical potential of a particle, as a result of this particle--hole symmetry~\cite{pathria1996statistical}.
The equilibrium supersaturation, $S$, is thus approximately equal to $\exp[\beta(\mu-2\epsilon)]$, which can be interpreted as the ratio $\rho/(1-\rho)$ at $u=2\epsilon$, where $\rho$ is the number density of particles.
We take this ratio as the definition of supersaturation $\tilde{S}$ in the FLEX-mapped effective equilibrium,
\begin{equation} \label{eq:FLEX_S}
    \tilde{S} \equiv \left[\dfrac{\rhoBT}{1-\rhoBT}\right]_{u=2\epsilon} = \dfrac{\zBT(2\epsilon;\; \Dmu)e^{-2\beta\epsilon}}{\zIT(2\epsilon;\; \Dmu)+1}.
\end{equation}
We then relate the FLEX supersaturation, $\tilde{S}$, to the ``thermodynamic'' driving force of the nonequilibrium phase transition, $\DPhi$, via $\tilde{S} \equiv \exp(\beta\DPhi)$, and we associate phase coexistence with $\tilde{S} = 1$.
The number density of particles in the vapor phase, $\rhov$, is taken to be the sum of the B and I-state number densities in the effective equilibrium at $u = 0$,
\begin{equation} \label{eq:FLEX_rhov}
\rhov \equiv \rhoBT(0;\;\Dmu)+\rhoIT(0;\;\Dmu).
\end{equation}
Here we assume that the vapor phase is sparsely populated with particles and so interactions among them are negligible.
Under a specified chemical drive at phase coexistence, $\Dmucoex$, Eqs.~(\ref{eq:FLEX_S}) and (\ref{eq:FLEX_rhov}) uniquely determine the effective fugacities and effective equilibrium of the open system.

\section{Application of FLEX to chemical-reaction models}
\label{app:FLEX-models}

Eqs.~(\ref{eq:FLEX_S}) and (\ref{eq:FLEX_rhov}) can be simplified to provide a qualitative understanding of the nonequilibrium phase behavior of the macroscopic liquid and vapor phases at coexistence.
For the kinetic models employed in this paper, the reaction rate $\kIB$ plateaus at low potential energies, either approaching $k^\circ$ in model I or $0$ in model II (\figref{fig:fig1}c).
The plateau allows us to approximately equate $\kIB(4\epsilon)$ and $\kIB(2\epsilon)$, meaning that the chemical-reaction kinetics in the liquid phase can be used to determine phase coexistence.
We further assume that the I state is much more stable than the B state so that $e^{\beta\Dfres} \gg 1$.
With these approximations, we can directly evaluate the coexistence conditions for both chemical-reaction models.

In the case of chemical-reaction model I, $\kIB$ approaches zero at $u = 0$ (\figref{fig:fig1}b), which leads to identical fugacities in the reservoir and in the FLEX approximation of the vapor phase.
Under this condition, the FLEX coexistence criterion, $\tilde{S} = 1$, leads to a self-consistent equation for $\Dfres$,
\begin{equation}
    e^{\beta\Dfres} = \rhov e^{-2\beta\epsilon} \left[ \dfrac{1+\kIB(4\epsilon)(1+e^{\beta\Dfres})}{1+\kIB(4\epsilon)(1+e^{\beta\Dfres})e^{\beta\Dmu}}\right].
    \label{eq:FLEX_simplified}
\end{equation}
Substituting \eqref{eq:FLEX_simplified} into \eqref{eq:df_FLEX} yields $\beta\Dfl = \ln(\rhov e^{-2\beta\epsilon})$, which remains constant along the coexistence line.
When $\rhov$ is much higher than the coexistence condition for the equilibrium two-state lattice-gas model, $\rhov e^{-2\beta\epsilon} \gg 1$, this prediction for $\Dfl$ agrees well with the FLEX result at equilibrium, $\beta\Dfres = \beta\Dfl = \beta\Df_\text{eq} = \ln\left[\rhov(e^{-2\beta\epsilon}+1)-1\right]$.

In the case of chemical-reaction model II, $\kIB$ becomes small near $u = 2\epsilon$, so that the FLEX fugacities become identical to the reservoir values at this value of $u$.
Applying this condition to Eqs.~(\ref{eq:FLEX_S}) and (\ref{eq:FLEX_rhov}) leads to a result that is similar to \eqref{eq:FLEX_simplified} in the case of model I,
\begin{align}
\begin{split}
    e^{\beta\Dfres} &= \rhov e^{-2\beta\epsilon} \left[\dfrac{1+\kIB(0)\left\{1+e^{\beta(\Dfres + \Dmu)}\right\}}{1+\kIB(0)(1+e^{\beta\Dfres})e^{\beta\Dmu}}\right] \\
    &\approx \rhov e^{-2\beta\epsilon},
    \label{eq:FLEX_simplified_model2}
\end{split}
\end{align}
where the approximation on the second line follows from the assumption $e^{\beta\Dfres} \gg 1$.
\eqref{eq:FLEX_simplified_model2} indicates that $\Dfres$ is nearly independent of $\Dmucoex$, which agrees with the results shown in \figref{fig:fig3}a.
Evaluating \eqref{eq:df_FLEX} leads to $\beta\Dfl \approx \beta\Dfres$, since $\kIB(4\epsilon)\approx0$ in this model.

\section{Predicting interfacial properties with FLEX}
\label{app:FLEX-interface}

We now apply FLEX to predict the nonequilibrium interfacial tension.
To this end, we focus on the effective free-energy cost of attaching a B-state particle to a flat liquid--vapor interface.
Since the free-energy cost of attaching a B-state adatom is $-\beta\epsilon$ in equilibrium, we propose that an appropriate effective-equilibrium model of a nonequilibrium interface should reproduce the relation $\rhoB = \exp(\beta\epsilon) / [1 + \exp(\beta\epsilon)]$ at a tagged lattice site on the interface.
We therefore invert this relationship to extract the effective dimensionless interaction strength, $\beta\tilde{\epsilon}$, of an adatom at the interface at steady state,
\begin{equation} \label{eq:FLEX_epsilon}
    \beta\tilde{\epsilon} \equiv \ln\left[\dfrac{\rhoBT}{1-\rhoBT}\right]_{u=\epsilon} = \ln\left[\dfrac{\zBT(\epsilon;\;\Dmucoex)}{\zIT(\epsilon;\;\Dmucoex)+1}\right]-\beta\epsilon,
\end{equation}
where $\tilde\rhoB$ is the FLEX approximation of the B-state number density at coexistence.
The notation $[\cdot]_{u=\epsilon}$ indicates that the tagged lattice site experiences a fixed local environment consisting of exactly one B-state particle.
Evaluating \eqref{eq:roughness_ratio} using this effective interaction strength predicts the deviation of the interfacial roughness from the equilibrium value within the FLEX framework.

From the effective dimensionless interaction strength at the interface, $\beta\tilde{\epsilon}$, we predict the nonequilibrium interfacial tension, $\sigma(\tilde{\epsilon})$, using the following analytical formula at equilibrium~\cite{shneidman1999analytical}:
\begin{equation}
\label{eq:sigma_analytical}
    \sigma(\tilde\epsilon) = \!\sqrt{\dfrac{4\tilde\epsilon\beta^{-2}}{\pi\chi(\beta)}\!\!\int_{\beta_c}^\beta \!\!\! K'\!\left(\!\dfrac{8[\cosh(\beta'\tilde\epsilon)\!-\!1]}{[\cosh(\beta'\tilde\epsilon)\!+\!1]^2}\!\right) \!\!\left[\!\dfrac{\cosh(\beta'\tilde\epsilon)\!-\!3}{\sinh(\beta'\tilde\epsilon)}\!\right] \!d\beta'}\!,
\end{equation}
where $K'$ is the elliptic integral of the first kind, $\chi(\beta) = [1-\sinh^{-4}({\beta\tilde{\epsilon}/2})]^{1/8}$, and $\beta_\text{c}$ is the inverse critical temperature given by $\beta_\text{c}|\tilde{\epsilon}| = 2\ln(1+\sqrt{2})$.
This prediction captures the general trend of the interfacial tension under nonequilibrium conditions with different chemical-reaction kinetics (\figref{fig:fig5}a).

\providecommand{\noopsort}[1]{}\providecommand{\singleletter}[1]{#1}%


\begin{thebibliography}{48}%
\makeatletter
\providecommand \@ifxundefined [1]{%
 \@ifx{#1\undefined}
}%
\providecommand \@ifnum [1]{%
 \ifnum #1\expandafter \@firstoftwo
 \else \expandafter \@secondoftwo
 \fi
}%
\providecommand \@ifx [1]{%
 \ifx #1\expandafter \@firstoftwo
 \else \expandafter \@secondoftwo
 \fi
}%
\providecommand \natexlab [1]{#1}%
\providecommand \enquote  [1]{``#1''}%
\providecommand \bibnamefont  [1]{#1}%
\providecommand \bibfnamefont [1]{#1}%
\providecommand \citenamefont [1]{#1}%
\providecommand \href@noop [0]{\@secondoftwo}%
\providecommand \href [0]{\begingroup \@sanitize@url \@href}%
\providecommand \@href[1]{\@@startlink{#1}\@@href}%
\providecommand \@@href[1]{\endgroup#1\@@endlink}%
\providecommand \@sanitize@url [0]{\catcode `\\12\catcode `\$12\catcode
  `\&12\catcode `\#12\catcode `\^12\catcode `\_12\catcode `\%12\relax}%
\providecommand \@@startlink[1]{}%
\providecommand \@@endlink[0]{}%
\providecommand \url  [0]{\begingroup\@sanitize@url \@url }%
\providecommand \@url [1]{\endgroup\@href {#1}{\urlprefix }}%
\providecommand \urlprefix  [0]{URL }%
\providecommand \Eprint [0]{\href }%
\providecommand \doibase [0]{https://doi.org/}%
\providecommand \selectlanguage [0]{\@gobble}%
\providecommand \bibinfo  [0]{\@secondoftwo}%
\providecommand \bibfield  [0]{\@secondoftwo}%
\providecommand \translation [1]{[#1]}%
\providecommand \BibitemOpen [0]{}%
\providecommand \bibitemStop [0]{}%
\providecommand \bibitemNoStop [0]{.\EOS\space}%
\providecommand \EOS [0]{\spacefactor3000\relax}%
\providecommand \BibitemShut  [1]{\csname bibitem#1\endcsname}%
\let\auto@bib@innerbib\@empty
\bibitem [{\citenamefont {Weber}\ \emph {et~al.}(2019)\citenamefont {Weber},
  \citenamefont {Zwicker}, \citenamefont {J{\"u}licher},\ and\ \citenamefont
  {Lee}}]{weber2019physics}%
  \BibitemOpen
  \bibfield  {author} {\bibinfo {author} {\bibfnamefont {C.~A.}\ \bibnamefont
  {Weber}}, \bibinfo {author} {\bibfnamefont {D.}~\bibnamefont {Zwicker}},
  \bibinfo {author} {\bibfnamefont {F.}~\bibnamefont {J{\"u}licher}},\ and\
  \bibinfo {author} {\bibfnamefont {C.~F.}\ \bibnamefont {Lee}},\ }\bibfield
  {title} {\enquote {\bibinfo {title} {Physics of active emulsions},}\ }\href
  {https://doi.org/10.1088/1361-6633/ab052b} {\bibfield  {journal} {\bibinfo
  {journal} {Rep. Prog. Phys.}\ }\textbf {\bibinfo {volume} {82}},\ \bibinfo
  {pages} {064601} (\bibinfo {year} {2019})}\BibitemShut {NoStop}%
\bibitem [{\citenamefont {Zwicker}(2022)}]{zwicker2022intertwined}%
  \BibitemOpen
  \bibfield  {author} {\bibinfo {author} {\bibfnamefont {D.}~\bibnamefont
  {Zwicker}},\ }\bibfield  {title} {\enquote {\bibinfo {title} {The intertwined
  physics of active chemical reactions and phase separation},}\ }\href
  {https://doi.org/https://doi.org/10.1016/j.cocis.2022.101606} {\bibfield
  {journal} {\bibinfo  {journal} {Curr. Opin. Coll. Inter. Sci.}\ }\textbf
  {\bibinfo {volume} {61}},\ \bibinfo {pages} {101606} (\bibinfo {year}
  {2022})}\BibitemShut {NoStop}%
\bibitem [{\citenamefont {Monahan}\ \emph {et~al.}(2017)\citenamefont
  {Monahan}, \citenamefont {Ryan}, \citenamefont {Janke}, \citenamefont
  {Burke}, \citenamefont {Rhoads}, \citenamefont {Zerze}, \citenamefont
  {O'Meally}, \citenamefont {Dignon}, \citenamefont {Conicella}, \citenamefont
  {Zheng}, \citenamefont {Best}, \citenamefont {Cole}, \citenamefont {Mittal},
  \citenamefont {Shewmaker},\ and\ \citenamefont
  {Fawzi}}]{monahan2017phosphorylation}%
  \BibitemOpen
  \bibfield  {author} {\bibinfo {author} {\bibfnamefont {Z.}~\bibnamefont
  {Monahan}}, \bibinfo {author} {\bibfnamefont {V.~H.}\ \bibnamefont {Ryan}},
  \bibinfo {author} {\bibfnamefont {A.~M.}\ \bibnamefont {Janke}}, \bibinfo
  {author} {\bibfnamefont {K.~A.}\ \bibnamefont {Burke}}, \bibinfo {author}
  {\bibfnamefont {S.~N.}\ \bibnamefont {Rhoads}}, \bibinfo {author}
  {\bibfnamefont {G.~H.}\ \bibnamefont {Zerze}}, \bibinfo {author}
  {\bibfnamefont {R.}~\bibnamefont {O'Meally}}, \bibinfo {author}
  {\bibfnamefont {G.~L.}\ \bibnamefont {Dignon}}, \bibinfo {author}
  {\bibfnamefont {A.~E.}\ \bibnamefont {Conicella}}, \bibinfo {author}
  {\bibfnamefont {W.}~\bibnamefont {Zheng}}, \bibinfo {author} {\bibfnamefont
  {R.~B.}\ \bibnamefont {Best}}, \bibinfo {author} {\bibfnamefont {R.~N.}\
  \bibnamefont {Cole}}, \bibinfo {author} {\bibfnamefont {J.}~\bibnamefont
  {Mittal}}, \bibinfo {author} {\bibfnamefont {F.}~\bibnamefont {Shewmaker}},\
  and\ \bibinfo {author} {\bibfnamefont {N.~L.}\ \bibnamefont {Fawzi}},\
  }\bibfield  {title} {\enquote {\bibinfo {title} {Phosphorylation of the {FUS}
  low-complexity domain disrupts phase separation, aggregation, and
  toxicity},}\ }\href {https://doi.org/https://doi.org/10.15252/embj.201696394}
  {\bibfield  {journal} {\bibinfo  {journal} {EMBO J.}\ }\textbf {\bibinfo
  {volume} {36}},\ \bibinfo {pages} {2951} (\bibinfo {year}
  {2017})}\BibitemShut {NoStop}%
\bibitem [{\citenamefont {Kim}\ \emph {et~al.}(2019)\citenamefont {Kim},
  \citenamefont {Tsang}, \citenamefont {Vernon}, \citenamefont {Sonenberg},
  \citenamefont {Kay},\ and\ \citenamefont
  {Forman-Kay}}]{kim2019phosphorylation}%
  \BibitemOpen
  \bibfield  {author} {\bibinfo {author} {\bibfnamefont {T.~H.}\ \bibnamefont
  {Kim}}, \bibinfo {author} {\bibfnamefont {B.}~\bibnamefont {Tsang}}, \bibinfo
  {author} {\bibfnamefont {R.~M.}\ \bibnamefont {Vernon}}, \bibinfo {author}
  {\bibfnamefont {N.}~\bibnamefont {Sonenberg}}, \bibinfo {author}
  {\bibfnamefont {L.~E.}\ \bibnamefont {Kay}},\ and\ \bibinfo {author}
  {\bibfnamefont {J.~D.}\ \bibnamefont {Forman-Kay}},\ }\bibfield  {title}
  {\enquote {\bibinfo {title} {Phospho-dependent phase separation of {FMRP} and
  {CAPRIN1} recapitulates regulation of translation and deadenylation},}\
  }\href {https://doi.org/10.1126/science.aax4240} {\bibfield  {journal}
  {\bibinfo  {journal} {Science}\ }\textbf {\bibinfo {volume} {365}},\ \bibinfo
  {pages} {825} (\bibinfo {year} {2019})}\BibitemShut {NoStop}%
\bibitem [{\citenamefont {Nosella}\ and\ \citenamefont
  {Forman-Kay}(2021)}]{nosella2021phosphorylation}%
  \BibitemOpen
  \bibfield  {author} {\bibinfo {author} {\bibfnamefont {M.~L.}\ \bibnamefont
  {Nosella}}\ and\ \bibinfo {author} {\bibfnamefont {J.~D.}\ \bibnamefont
  {Forman-Kay}},\ }\bibfield  {title} {\enquote {\bibinfo {title}
  {Phosphorylation-dependent regulation of messenger {RNA} transcription,
  processing and translation within biomolecular condensates},}\ }\href
  {https://doi.org/https://doi.org/10.1016/j.ceb.2020.12.007} {\bibfield
  {journal} {\bibinfo  {journal} {Curr. Opin. Cell Biol.}\ }\textbf {\bibinfo
  {volume} {69}},\ \bibinfo {pages} {30} (\bibinfo {year} {2021})}\BibitemShut
  {NoStop}%
\bibitem [{\citenamefont {Berry}, \citenamefont {Brangwynne},\ and\
  \citenamefont {Haataja}(2018)}]{berry2018physical}%
  \BibitemOpen
  \bibfield  {author} {\bibinfo {author} {\bibfnamefont {J.}~\bibnamefont
  {Berry}}, \bibinfo {author} {\bibfnamefont {C.~P.}\ \bibnamefont
  {Brangwynne}},\ and\ \bibinfo {author} {\bibfnamefont {M.}~\bibnamefont
  {Haataja}},\ }\bibfield  {title} {\enquote {\bibinfo {title} {Physical
  principles of intracellular organization via active and passive phase
  transitions},}\ }\href {https://doi.org/10.1088/1361-6633/aaa61e} {\bibfield
  {journal} {\bibinfo  {journal} {Rep. Prog. Phys.}\ }\textbf {\bibinfo
  {volume} {81}},\ \bibinfo {pages} {046601} (\bibinfo {year}
  {2018})}\BibitemShut {NoStop}%
\bibitem [{\citenamefont {Zwicker}, \citenamefont {Hyman},\ and\ \citenamefont
  {J{\"u}licher}(2015)}]{zwicker2015suppression}%
  \BibitemOpen
  \bibfield  {author} {\bibinfo {author} {\bibfnamefont {D.}~\bibnamefont
  {Zwicker}}, \bibinfo {author} {\bibfnamefont {A.~A.}\ \bibnamefont {Hyman}},\
  and\ \bibinfo {author} {\bibfnamefont {F.}~\bibnamefont {J{\"u}licher}},\
  }\bibfield  {title} {\enquote {\bibinfo {title} {Suppression of {O}stwald
  ripening in active emulsions},}\ }\href
  {https://doi.org/10.1103/PhysRevE.92.012317} {\bibfield  {journal} {\bibinfo
  {journal} {Phys. Rev. E}\ }\textbf {\bibinfo {volume} {92}},\ \bibinfo
  {pages} {012317} (\bibinfo {year} {2015})}\BibitemShut {NoStop}%
\bibitem [{\citenamefont {Wurtz}\ and\ \citenamefont
  {Lee}(2018)}]{wurtz2018chemical}%
  \BibitemOpen
  \bibfield  {author} {\bibinfo {author} {\bibfnamefont {J.~D.}\ \bibnamefont
  {Wurtz}}\ and\ \bibinfo {author} {\bibfnamefont {C.~F.}\ \bibnamefont
  {Lee}},\ }\bibfield  {title} {\enquote {\bibinfo {title}
  {Chemical-reaction-controlled phase separated drops: {F}ormation, size
  selection, and coarsening},}\ }\href
  {https://doi.org/10.1103/PhysRevLett.120.078102} {\bibfield  {journal}
  {\bibinfo  {journal} {Phys. Rev. Lett.}\ }\textbf {\bibinfo {volume} {120}},\
  \bibinfo {pages} {078102} (\bibinfo {year} {2018})}\BibitemShut {NoStop}%
\bibitem [{\citenamefont {Kirschbaum}\ and\ \citenamefont
  {Zwicker}(2021)}]{kirschbaum2021controlling}%
  \BibitemOpen
  \bibfield  {author} {\bibinfo {author} {\bibfnamefont {J.}~\bibnamefont
  {Kirschbaum}}\ and\ \bibinfo {author} {\bibfnamefont {D.}~\bibnamefont
  {Zwicker}},\ }\bibfield  {title} {\enquote {\bibinfo {title} {Controlling
  biomolecular condensates via chemical reactions},}\ }\href
  {https://doi.org/10.1098/rsif.2021.0255} {\bibfield  {journal} {\bibinfo
  {journal} {J. R. Soc. Interface}\ }\textbf {\bibinfo {volume} {18}},\
  \bibinfo {pages} {20210255} (\bibinfo {year} {2021})}\BibitemShut {NoStop}%
\bibitem [{\citenamefont {Tena-Solsona}\ \emph {et~al.}(2021)\citenamefont
  {Tena-Solsona}, \citenamefont {Janssen}, \citenamefont {Wanzke},
  \citenamefont {Schnitter}, \citenamefont {Park}, \citenamefont {Rie{\ss}},
  \citenamefont {Gibbs}, \citenamefont {Weber},\ and\ \citenamefont
  {Boekhoven}}]{Tena-Solsona2021ripeningaccelerated}%
  \BibitemOpen
  \bibfield  {author} {\bibinfo {author} {\bibfnamefont {M.}~\bibnamefont
  {Tena-Solsona}}, \bibinfo {author} {\bibfnamefont {J.}~\bibnamefont
  {Janssen}}, \bibinfo {author} {\bibfnamefont {C.}~\bibnamefont {Wanzke}},
  \bibinfo {author} {\bibfnamefont {F.}~\bibnamefont {Schnitter}}, \bibinfo
  {author} {\bibfnamefont {H.}~\bibnamefont {Park}}, \bibinfo {author}
  {\bibfnamefont {B.}~\bibnamefont {Rie{\ss}}}, \bibinfo {author}
  {\bibfnamefont {J.~M.}\ \bibnamefont {Gibbs}}, \bibinfo {author}
  {\bibfnamefont {C.~A.}\ \bibnamefont {Weber}},\ and\ \bibinfo {author}
  {\bibfnamefont {J.}~\bibnamefont {Boekhoven}},\ }\bibfield  {title} {\enquote
  {\bibinfo {title} {Accelerated ripening in chemically fueled emulsions},}\
  }\href {https://doi.org/https://doi.org/10.1002/syst.202000034} {\bibfield
  {journal} {\bibinfo  {journal} {Chem. Sys. Chem.}\ }\textbf {\bibinfo
  {volume} {3}},\ \bibinfo {pages} {e2000034} (\bibinfo {year}
  {2021})}\BibitemShut {NoStop}%
\bibitem [{\citenamefont {Zwicker}\ \emph {et~al.}(2017)\citenamefont
  {Zwicker}, \citenamefont {Seyboldt}, \citenamefont {Weber}, \citenamefont
  {Hyman},\ and\ \citenamefont {J{\"u}licher}}]{zwicker2017growth}%
  \BibitemOpen
  \bibfield  {author} {\bibinfo {author} {\bibfnamefont {D.}~\bibnamefont
  {Zwicker}}, \bibinfo {author} {\bibfnamefont {R.}~\bibnamefont {Seyboldt}},
  \bibinfo {author} {\bibfnamefont {C.~A.}\ \bibnamefont {Weber}}, \bibinfo
  {author} {\bibfnamefont {A.~A.}\ \bibnamefont {Hyman}},\ and\ \bibinfo
  {author} {\bibfnamefont {F.}~\bibnamefont {J{\"u}licher}},\ }\bibfield
  {title} {\enquote {\bibinfo {title} {Growth and division of active droplets
  provides a model for protocells},}\ }\href
  {https://doi.org/10.1038/nphys3984} {\bibfield  {journal} {\bibinfo
  {journal} {Nature Phys.}\ }\textbf {\bibinfo {volume} {13}},\ \bibinfo
  {pages} {408} (\bibinfo {year} {2017})}\BibitemShut {NoStop}%
\bibitem [{\citenamefont {Demarchi}\ \emph {et~al.}(2023)\citenamefont
  {Demarchi}, \citenamefont {Goychuk}, \citenamefont {Maryshev},\ and\
  \citenamefont {Frey}}]{Demarchi2023Selfpropulsion}%
  \BibitemOpen
  \bibfield  {author} {\bibinfo {author} {\bibfnamefont {L.}~\bibnamefont
  {Demarchi}}, \bibinfo {author} {\bibfnamefont {A.}~\bibnamefont {Goychuk}},
  \bibinfo {author} {\bibfnamefont {I.}~\bibnamefont {Maryshev}},\ and\
  \bibinfo {author} {\bibfnamefont {E.}~\bibnamefont {Frey}},\ }\bibfield
  {title} {\enquote {\bibinfo {title} {Enzyme-enriched condensates show
  self-propulsion, positioning, and coexistence},}\ }\href
  {https://doi.org/10.1103/PhysRevLett.130.128401} {\bibfield  {journal}
  {\bibinfo  {journal} {Phys. Rev. Lett.}\ }\textbf {\bibinfo {volume} {130}},\
  \bibinfo {pages} {128401} (\bibinfo {year} {2023})}\BibitemShut {NoStop}%
\bibitem [{\citenamefont {Bauermann}\ \emph {et~al.}(2022)\citenamefont
  {Bauermann}, \citenamefont {Laha}, \citenamefont {McCall}, \citenamefont
  {J{\"u}licher},\ and\ \citenamefont {Weber}}]{bauermann2022chemical}%
  \BibitemOpen
  \bibfield  {author} {\bibinfo {author} {\bibfnamefont {J.}~\bibnamefont
  {Bauermann}}, \bibinfo {author} {\bibfnamefont {S.}~\bibnamefont {Laha}},
  \bibinfo {author} {\bibfnamefont {P.~M.}\ \bibnamefont {McCall}}, \bibinfo
  {author} {\bibfnamefont {F.}~\bibnamefont {J{\"u}licher}},\ and\ \bibinfo
  {author} {\bibfnamefont {C.~A.}\ \bibnamefont {Weber}},\ }\bibfield  {title}
  {\enquote {\bibinfo {title} {Chemical kinetics and mass action in coexisting
  phases},}\ }\href {https://doi.org/https://doi.org/10.1021/jacs.2c06265}
  {\bibfield  {journal} {\bibinfo  {journal} {J. Am. Chem. Soc.}\ }\textbf
  {\bibinfo {volume} {144}},\ \bibinfo {pages} {19294} (\bibinfo {year}
  {2022})}\BibitemShut {NoStop}%
\bibitem [{\citenamefont {Donau}\ \emph {et~al.}(2022)\citenamefont {Donau},
  \citenamefont {Sp{\"a}th}, \citenamefont {Stasi}, \citenamefont {Bergmann},\
  and\ \citenamefont {Boekhoven}}]{Donau2022multiphasic}%
  \BibitemOpen
  \bibfield  {author} {\bibinfo {author} {\bibfnamefont {C.}~\bibnamefont
  {Donau}}, \bibinfo {author} {\bibfnamefont {F.}~\bibnamefont {Sp{\"a}th}},
  \bibinfo {author} {\bibfnamefont {M.}~\bibnamefont {Stasi}}, \bibinfo
  {author} {\bibfnamefont {A.~M.}\ \bibnamefont {Bergmann}},\ and\ \bibinfo
  {author} {\bibfnamefont {J.}~\bibnamefont {Boekhoven}},\ }\bibfield  {title}
  {\enquote {\bibinfo {title} {Phase transitions in chemically fueled,
  multiphase complex coacervate droplets},}\ }\href
  {https://doi.org/https://doi.org/10.1002/anie.202211905} {\bibfield
  {journal} {\bibinfo  {journal} {Angew. Chem. Int. Ed.}\ }\textbf {\bibinfo
  {volume} {61}},\ \bibinfo {pages} {e202211905} (\bibinfo {year}
  {2022})}\BibitemShut {NoStop}%
\bibitem [{\citenamefont {Cho}\ and\ \citenamefont
  {Jacobs}(2023)}]{cho2023nucleation}%
  \BibitemOpen
  \bibfield  {author} {\bibinfo {author} {\bibfnamefont {Y.}~\bibnamefont
  {Cho}}\ and\ \bibinfo {author} {\bibfnamefont {W.~M.}\ \bibnamefont
  {Jacobs}},\ }\bibfield  {title} {\enquote {\bibinfo {title} {Tuning
  nucleation kinetics via nonequilibrium chemical reactions},}\ }\href
  {https://doi.org/10.1103/PhysRevLett.130.128203} {\bibfield  {journal}
  {\bibinfo  {journal} {Phys. Rev. Lett.}\ }\textbf {\bibinfo {volume} {130}},\
  \bibinfo {pages} {128203} (\bibinfo {year} {2023})}\BibitemShut {NoStop}%
\bibitem [{\citenamefont {Cates}\ and\ \citenamefont
  {Nardini}(2023)}]{cates2023nucleation}%
  \BibitemOpen
  \bibfield  {author} {\bibinfo {author} {\bibfnamefont {M.~E.}\ \bibnamefont
  {Cates}}\ and\ \bibinfo {author} {\bibfnamefont {C.}~\bibnamefont
  {Nardini}},\ }\bibfield  {title} {\enquote {\bibinfo {title} {Classical
  nucleation theory for active fluid phase separation},}\ }\href
  {https://doi.org/10.1103/PhysRevLett.130.098203} {\bibfield  {journal}
  {\bibinfo  {journal} {Phys. Rev. Lett.}\ }\textbf {\bibinfo {volume} {130}},\
  \bibinfo {pages} {098203} (\bibinfo {year} {2023})}\BibitemShut {NoStop}%
\bibitem [{\citenamefont {Ziethen}, \citenamefont {Kirschbaum},\ and\
  \citenamefont {Zwicker}(2023)}]{Ziethen2023nucleation}%
  \BibitemOpen
  \bibfield  {author} {\bibinfo {author} {\bibfnamefont {N.}~\bibnamefont
  {Ziethen}}, \bibinfo {author} {\bibfnamefont {J.}~\bibnamefont
  {Kirschbaum}},\ and\ \bibinfo {author} {\bibfnamefont {D.}~\bibnamefont
  {Zwicker}},\ }\bibfield  {title} {\enquote {\bibinfo {title} {Nucleation of
  chemically active droplets},}\ }\href
  {https://doi.org/10.1103/PhysRevLett.130.248201} {\bibfield  {journal}
  {\bibinfo  {journal} {Phys. Rev. Lett.}\ }\textbf {\bibinfo {volume} {130}},\
  \bibinfo {pages} {248201} (\bibinfo {year} {2023})}\BibitemShut {NoStop}%
\bibitem [{\citenamefont {Nott}\ \emph {et~al.}(2015)\citenamefont {Nott},
  \citenamefont {Petsalaki}, \citenamefont {Farber}, \citenamefont {Jervis},
  \citenamefont {Fussner}, \citenamefont {Plochowietz}, \citenamefont {Craggs},
  \citenamefont {Bazett-Jones}, \citenamefont {Pawson}, \citenamefont
  {Forman-Kay},\ and\ \citenamefont {Baldwin}}]{Nott2015PhaseTransition}%
  \BibitemOpen
  \bibfield  {author} {\bibinfo {author} {\bibfnamefont {T.~J.}\ \bibnamefont
  {Nott}}, \bibinfo {author} {\bibfnamefont {E.}~\bibnamefont {Petsalaki}},
  \bibinfo {author} {\bibfnamefont {P.}~\bibnamefont {Farber}}, \bibinfo
  {author} {\bibfnamefont {D.}~\bibnamefont {Jervis}}, \bibinfo {author}
  {\bibfnamefont {E.}~\bibnamefont {Fussner}}, \bibinfo {author} {\bibfnamefont
  {A.}~\bibnamefont {Plochowietz}}, \bibinfo {author} {\bibfnamefont {T.~D.}\
  \bibnamefont {Craggs}}, \bibinfo {author} {\bibfnamefont {D.~P.}\
  \bibnamefont {Bazett-Jones}}, \bibinfo {author} {\bibfnamefont
  {T.}~\bibnamefont {Pawson}}, \bibinfo {author} {\bibfnamefont {J.~D.}\
  \bibnamefont {Forman-Kay}},\ and\ \bibinfo {author} {\bibfnamefont {A.~J.}\
  \bibnamefont {Baldwin}},\ }\bibfield  {title} {\enquote {\bibinfo {title}
  {Phase transition of a disordered nuage protein generates environmentally
  responsive membraneless organelles},}\ }\href
  {https://doi.org/10.1016/j.molcel.2015.01.013} {\bibfield  {journal}
  {\bibinfo  {journal} {Mol. Cell}\ }\textbf {\bibinfo {volume} {57}},\
  \bibinfo {pages} {936} (\bibinfo {year} {2015})}\BibitemShut {NoStop}%
\bibitem [{\citenamefont {S{\"o}ding}\ \emph {et~al.}(2020)\citenamefont
  {S{\"o}ding}, \citenamefont {Zwicker}, \citenamefont {Sohrabi-Jahromi},
  \citenamefont {Boehning},\ and\ \citenamefont
  {Kirschbaum}}]{soding2020mechanisms}%
  \BibitemOpen
  \bibfield  {author} {\bibinfo {author} {\bibfnamefont {J.}~\bibnamefont
  {S{\"o}ding}}, \bibinfo {author} {\bibfnamefont {D.}~\bibnamefont {Zwicker}},
  \bibinfo {author} {\bibfnamefont {S.}~\bibnamefont {Sohrabi-Jahromi}},
  \bibinfo {author} {\bibfnamefont {M.}~\bibnamefont {Boehning}},\ and\
  \bibinfo {author} {\bibfnamefont {J.}~\bibnamefont {Kirschbaum}},\ }\bibfield
   {title} {\enquote {\bibinfo {title} {Mechanisms for active regulation of
  biomolecular condensates},}\ }\href
  {https://doi.org/10.1016/j.tcb.2019.10.006} {\bibfield  {journal} {\bibinfo
  {journal} {Trends Cell Biol.}\ }\textbf {\bibinfo {volume} {30}},\ \bibinfo
  {pages} {4} (\bibinfo {year} {2020})}\BibitemShut {NoStop}%
\bibitem [{\citenamefont {Besse}\ \emph {et~al.}(2023)\citenamefont {Besse},
  \citenamefont {Fausti}, \citenamefont {Cates}, \citenamefont {Delamotte},\
  and\ \citenamefont {Nardini}}]{besse2023interface}%
  \BibitemOpen
  \bibfield  {author} {\bibinfo {author} {\bibfnamefont {M.}~\bibnamefont
  {Besse}}, \bibinfo {author} {\bibfnamefont {G.}~\bibnamefont {Fausti}},
  \bibinfo {author} {\bibfnamefont {M.~E.}\ \bibnamefont {Cates}}, \bibinfo
  {author} {\bibfnamefont {B.}~\bibnamefont {Delamotte}},\ and\ \bibinfo
  {author} {\bibfnamefont {C.}~\bibnamefont {Nardini}},\ }\bibfield  {title}
  {\enquote {\bibinfo {title} {Interface roughening in nonequilibrium
  phase-separated systems},}\ }\href
  {https://doi.org/https://doi.org/10.1103/PhysRevLett.130.187102} {\bibfield
  {journal} {\bibinfo  {journal} {Phys. Rev. Lett.}\ }\textbf {\bibinfo
  {volume} {130}},\ \bibinfo {pages} {187102} (\bibinfo {year}
  {2023})}\BibitemShut {NoStop}%
\bibitem [{\citenamefont {O'Flynn}\ and\ \citenamefont
  {Mittag}(2021)}]{oflynn2021role}%
  \BibitemOpen
  \bibfield  {author} {\bibinfo {author} {\bibfnamefont {B.~G.}\ \bibnamefont
  {O'Flynn}}\ and\ \bibinfo {author} {\bibfnamefont {T.}~\bibnamefont
  {Mittag}},\ }\bibfield  {title} {\enquote {\bibinfo {title} {The role of
  liquid--liquid phase separation in regulating enzyme activity},}\ }\href
  {https://doi.org/10.1016/j.ceb.2020.12.012} {\bibfield  {journal} {\bibinfo
  {journal} {Curr. Opin. Cell Biol.}\ }\textbf {\bibinfo {volume} {69}},\
  \bibinfo {pages} {70} (\bibinfo {year} {2021})}\BibitemShut {NoStop}%
\bibitem [{\citenamefont {Schuster}\ \emph {et~al.}(2021)\citenamefont
  {Schuster}, \citenamefont {Regy}, \citenamefont {Dolan}, \citenamefont
  {Kanchi~Ranganath}, \citenamefont {Jovic}, \citenamefont {Khare},
  \citenamefont {Shi},\ and\ \citenamefont
  {Mittal}}]{Schuster2021biocondensate_review}%
  \BibitemOpen
  \bibfield  {author} {\bibinfo {author} {\bibfnamefont {B.~S.}\ \bibnamefont
  {Schuster}}, \bibinfo {author} {\bibfnamefont {R.~M.}\ \bibnamefont {Regy}},
  \bibinfo {author} {\bibfnamefont {E.~M.}\ \bibnamefont {Dolan}}, \bibinfo
  {author} {\bibfnamefont {A.}~\bibnamefont {Kanchi~Ranganath}}, \bibinfo
  {author} {\bibfnamefont {N.}~\bibnamefont {Jovic}}, \bibinfo {author}
  {\bibfnamefont {S.~D.}\ \bibnamefont {Khare}}, \bibinfo {author}
  {\bibfnamefont {Z.}~\bibnamefont {Shi}},\ and\ \bibinfo {author}
  {\bibfnamefont {J.}~\bibnamefont {Mittal}},\ }\bibfield  {title} {\enquote
  {\bibinfo {title} {Biomolecular condensates: Sequence determinants of phase
  separation, microstructural organization, enzymatic activity, and material
  properties},}\ }\href {https://doi.org/10.1021/acs.jpcb.0c11606} {\bibfield
  {journal} {\bibinfo  {journal} {J. Phys. Chem. B}\ }\textbf {\bibinfo
  {volume} {125}},\ \bibinfo {pages} {3441} (\bibinfo {year}
  {2021})}\BibitemShut {NoStop}%
\bibitem [{\citenamefont {Seifert}(2012)}]{seifert2012stochastic}%
  \BibitemOpen
  \bibfield  {author} {\bibinfo {author} {\bibfnamefont {U.}~\bibnamefont
  {Seifert}},\ }\bibfield  {title} {\enquote {\bibinfo {title} {Stochastic
  thermodynamics, fluctuation theorems and molecular machines},}\ }\href
  {https://doi.org/10.1088/0034-4885/75/12/126001} {\bibfield  {journal}
  {\bibinfo  {journal} {Rep. Prog. Phys.}\ }\textbf {\bibinfo {volume} {75}},\
  \bibinfo {pages} {126001} (\bibinfo {year} {2012})}\BibitemShut {NoStop}%
\bibitem [{\citenamefont {Van~den Broeck}\ and\ \citenamefont
  {Esposito}(2015)}]{van2015ensemble}%
  \BibitemOpen
  \bibfield  {author} {\bibinfo {author} {\bibfnamefont {C.}~\bibnamefont
  {Van~den Broeck}}\ and\ \bibinfo {author} {\bibfnamefont {M.}~\bibnamefont
  {Esposito}},\ }\bibfield  {title} {\enquote {\bibinfo {title} {Ensemble and
  trajectory thermodynamics: {A} brief introduction},}\ }\href
  {https://doi.org/10.1016/j.physa.2014.04.035} {\bibfield  {journal} {\bibinfo
   {journal} {Physica A}\ }\textbf {\bibinfo {volume} {418}},\ \bibinfo {pages}
  {6} (\bibinfo {year} {2015})}\BibitemShut {NoStop}%
\bibitem [{\citenamefont {Wilding}(1995)}]{wilding1995critical}%
  \BibitemOpen
  \bibfield  {author} {\bibinfo {author} {\bibfnamefont {N.~B.}\ \bibnamefont
  {Wilding}},\ }\bibfield  {title} {\enquote {\bibinfo {title} {Critical-point
  and coexistence-curve properties of the {L}ennard-{J}ones fluid: {A}
  finite-size scaling study},}\ }\href
  {https://doi.org/10.1103/PhysRevE.52.602} {\bibfield  {journal} {\bibinfo
  {journal} {Phys. Rev. E}\ }\textbf {\bibinfo {volume} {52}},\ \bibinfo
  {pages} {602} (\bibinfo {year} {1995})}\BibitemShut {NoStop}%
\bibitem [{\citenamefont {Gillespie}(2007)}]{gillespie2007stochastic}%
  \BibitemOpen
  \bibfield  {author} {\bibinfo {author} {\bibfnamefont {D.~T.}\ \bibnamefont
  {Gillespie}},\ }\bibfield  {title} {\enquote {\bibinfo {title} {Stochastic
  simulation of chemical kinetics},}\ }\href
  {https://doi.org/10.1146/annurev.physchem.58.032806.104637} {\bibfield
  {journal} {\bibinfo  {journal} {Annu. Rev. Phys. Chem}\ }\textbf {\bibinfo
  {volume} {58}},\ \bibinfo {pages} {35} (\bibinfo {year} {2007})}\BibitemShut
  {NoStop}%
\bibitem [{\citenamefont {Pathria}(1996)}]{pathria1996statistical}%
  \BibitemOpen
  \bibfield  {author} {\bibinfo {author} {\bibfnamefont {R.~K.}\ \bibnamefont
  {Pathria}},\ }\href@noop {} {\emph {\bibinfo {title} {Statistical
  mechanics}}},\ \bibinfo {edition} {2nd}\ ed.\ (\bibinfo  {publisher}
  {Butterworth-Heinemann, Oxford},\ \bibinfo {year} {1996})\BibitemShut
  {NoStop}%
\bibitem [{\citenamefont {Chandler}(1987)}]{chandler1987introduction}%
  \BibitemOpen
  \bibfield  {author} {\bibinfo {author} {\bibfnamefont {D.}~\bibnamefont
  {Chandler}},\ }\href@noop {} {\emph {\bibinfo {title} {Introduction to modern
  statistical mechanics}}}\ (\bibinfo  {publisher} {Oxford University Press,
  New York},\ \bibinfo {year} {1987})\BibitemShut {NoStop}%
\bibitem [{\citenamefont {Hansen}\ and\ \citenamefont
  {McDonald}(2013)}]{hansen2013simpleliquids}%
  \BibitemOpen
  \bibfield  {author} {\bibinfo {author} {\bibfnamefont {J.-P.}\ \bibnamefont
  {Hansen}}\ and\ \bibinfo {author} {\bibfnamefont {I.~R.}\ \bibnamefont
  {McDonald}},\ }\href@noop {} {\emph {\bibinfo {title} {Theory of simple
  liquids: with applications to soft matter}}},\ \bibinfo {edition} {4th}\ ed.\
  (\bibinfo  {publisher} {Academic press, Oxford},\ \bibinfo {year}
  {2013})\BibitemShut {NoStop}%
\bibitem [{\citenamefont {Solon}\ \emph {et~al.}(2015)\citenamefont {Solon},
  \citenamefont {Fily}, \citenamefont {Baskaran}, \citenamefont {Cates},
  \citenamefont {Kafri}, \citenamefont {Kardar},\ and\ \citenamefont
  {Tailleur}}]{solon2015pressure}%
  \BibitemOpen
  \bibfield  {author} {\bibinfo {author} {\bibfnamefont {A.~P.}\ \bibnamefont
  {Solon}}, \bibinfo {author} {\bibfnamefont {Y.}~\bibnamefont {Fily}},
  \bibinfo {author} {\bibfnamefont {A.}~\bibnamefont {Baskaran}}, \bibinfo
  {author} {\bibfnamefont {M.~E.}\ \bibnamefont {Cates}}, \bibinfo {author}
  {\bibfnamefont {Y.}~\bibnamefont {Kafri}}, \bibinfo {author} {\bibfnamefont
  {M.}~\bibnamefont {Kardar}},\ and\ \bibinfo {author} {\bibfnamefont
  {J.}~\bibnamefont {Tailleur}},\ }\bibfield  {title} {\enquote {\bibinfo
  {title} {Pressure is not a state function for generic active fluids},}\
  }\href {https://doi.org/https://doi.org/10.1038/nphys3377} {\bibfield
  {journal} {\bibinfo  {journal} {Nature Phys.}\ }\textbf {\bibinfo {volume}
  {11}},\ \bibinfo {pages} {673} (\bibinfo {year} {2015})}\BibitemShut
  {NoStop}%
\bibitem [{\citenamefont {Omar}\ \emph {et~al.}(2023)\citenamefont {Omar},
  \citenamefont {Row}, \citenamefont {Mallory},\ and\ \citenamefont
  {Brady}}]{brady2023mechanical}%
  \BibitemOpen
  \bibfield  {author} {\bibinfo {author} {\bibfnamefont {A.~K.}\ \bibnamefont
  {Omar}}, \bibinfo {author} {\bibfnamefont {H.}~\bibnamefont {Row}}, \bibinfo
  {author} {\bibfnamefont {S.~A.}\ \bibnamefont {Mallory}},\ and\ \bibinfo
  {author} {\bibfnamefont {J.~F.}\ \bibnamefont {Brady}},\ }\bibfield  {title}
  {\enquote {\bibinfo {title} {Mechanical theory of nonequilibrium coexistence
  and motility-induced phase separation},}\ }\href
  {https://doi.org/10.1073/pnas.2219900120} {\bibfield  {journal} {\bibinfo
  {journal} {Proc. Natl. Acad. Sci. USA}\ }\textbf {\bibinfo {volume} {120}},\
  \bibinfo {pages} {e2219900120} (\bibinfo {year} {2023})}\BibitemShut
  {NoStop}%
\bibitem [{\citenamefont {Warmflash}, \citenamefont {Bhimalapuram},\ and\
  \citenamefont {Dinner}(2007)}]{warmflash2007umbrella}%
  \BibitemOpen
  \bibfield  {author} {\bibinfo {author} {\bibfnamefont {A.}~\bibnamefont
  {Warmflash}}, \bibinfo {author} {\bibfnamefont {P.}~\bibnamefont
  {Bhimalapuram}},\ and\ \bibinfo {author} {\bibfnamefont {A.~R.}\ \bibnamefont
  {Dinner}},\ }\bibfield  {title} {\enquote {\bibinfo {title} {Umbrella
  sampling for nonequilibrium processes},}\ }\href
  {https://doi.org/10.1063/1.2784118} {\bibfield  {journal} {\bibinfo
  {journal} {J. Chem. Phys.}\ }\textbf {\bibinfo {volume} {127}},\ \bibinfo
  {pages} {154112} (\bibinfo {year} {2007})}\BibitemShut {NoStop}%
\bibitem [{\citenamefont {Wilding}\ and\ \citenamefont
  {Bruce}(1992)}]{Wilding1992critical}%
  \BibitemOpen
  \bibfield  {author} {\bibinfo {author} {\bibfnamefont {N.~B.}\ \bibnamefont
  {Wilding}}\ and\ \bibinfo {author} {\bibfnamefont {A.~D.}\ \bibnamefont
  {Bruce}},\ }\bibfield  {title} {\enquote {\bibinfo {title} {Density
  fluctuations and field mixing in the critical fluid},}\ }\href
  {https://doi.org/10.1088/0953-8984/4/12/008} {\bibfield  {journal} {\bibinfo
  {journal} {J. Phys.: Condens. Matter}\ }\textbf {\bibinfo {volume} {4}},\
  \bibinfo {pages} {3087} (\bibinfo {year} {1992})}\BibitemShut {NoStop}%
\bibitem [{\citenamefont {Noya}, \citenamefont {Vega},\ and\ \citenamefont
  {de~Miguel}(2008)}]{noya2008}%
  \BibitemOpen
  \bibfield  {author} {\bibinfo {author} {\bibfnamefont {E.~G.}\ \bibnamefont
  {Noya}}, \bibinfo {author} {\bibfnamefont {C.}~\bibnamefont {Vega}},\ and\
  \bibinfo {author} {\bibfnamefont {E.}~\bibnamefont {de~Miguel}},\ }\bibfield
  {title} {\enquote {\bibinfo {title} {Determination of the melting point of
  hard spheres from direct coexistence simulation methods},}\ }\href
  {https://doi.org/10.1063/1.2901172} {\bibfield  {journal} {\bibinfo
  {journal} {J. Chem. Phys.}\ }\textbf {\bibinfo {volume} {128}},\ \bibinfo
  {pages} {154507} (\bibinfo {year} {2008})}\BibitemShut {NoStop}%
\bibitem [{\citenamefont {Christensen}, \citenamefont {Elder},\ and\
  \citenamefont {Fogedby}(1996)}]{Christensen1996coulombic}%
  \BibitemOpen
  \bibfield  {author} {\bibinfo {author} {\bibfnamefont {J.~J.}\ \bibnamefont
  {Christensen}}, \bibinfo {author} {\bibfnamefont {K.}~\bibnamefont {Elder}},\
  and\ \bibinfo {author} {\bibfnamefont {H.~C.}\ \bibnamefont {Fogedby}},\
  }\bibfield  {title} {\enquote {\bibinfo {title} {Phase segregation dynamics
  of a chemically reactive binary mixture},}\ }\href
  {https://doi.org/10.1103/PhysRevE.54.R2212} {\bibfield  {journal} {\bibinfo
  {journal} {Phys. Rev. E}\ }\textbf {\bibinfo {volume} {54}},\ \bibinfo
  {pages} {R2212} (\bibinfo {year} {1996})}\BibitemShut {NoStop}%
\bibitem [{\citenamefont {Allen}\ and\ \citenamefont
  {Tildesley}(2017)}]{Allen2017liquid}%
  \BibitemOpen
  \bibfield  {author} {\bibinfo {author} {\bibfnamefont {M.~P.}\ \bibnamefont
  {Allen}}\ and\ \bibinfo {author} {\bibfnamefont {D.~J.}\ \bibnamefont
  {Tildesley}},\ }\href@noop {} {\emph {\bibinfo {title} {Computer simulation
  of liquids}}},\ \bibinfo {edition} {2nd}\ ed.\ (\bibinfo  {publisher} {Oxford
  University Press, New York},\ \bibinfo {year} {2017})\BibitemShut {NoStop}%
\bibitem [{\citenamefont {Oxtoby}(1992)}]{oxtoby1992homogeneous}%
  \BibitemOpen
  \bibfield  {author} {\bibinfo {author} {\bibfnamefont {D.~W.}\ \bibnamefont
  {Oxtoby}},\ }\bibfield  {title} {\enquote {\bibinfo {title} {Homogeneous
  nucleation: {T}heory and experiment},}\ }\href
  {https://doi.org/10.1088/0953-8984/4/38/001} {\bibfield  {journal} {\bibinfo
  {journal} {J. Phys.: Condens. Matter}\ }\textbf {\bibinfo {volume} {4}},\
  \bibinfo {pages} {7627} (\bibinfo {year} {1992})}\BibitemShut {NoStop}%
\bibitem [{\citenamefont {Ryu}\ and\ \citenamefont
  {Cai}(2010)}]{ryu2010validity}%
  \BibitemOpen
  \bibfield  {author} {\bibinfo {author} {\bibfnamefont {S.}~\bibnamefont
  {Ryu}}\ and\ \bibinfo {author} {\bibfnamefont {W.}~\bibnamefont {Cai}},\
  }\bibfield  {title} {\enquote {\bibinfo {title} {Validity of classical
  nucleation theory for {I}sing models},}\ }\href
  {https://doi.org/10.1103/PhysRevE.81.030601} {\bibfield  {journal} {\bibinfo
  {journal} {Phys. Rev. E}\ }\textbf {\bibinfo {volume} {81}},\ \bibinfo
  {pages} {030601} (\bibinfo {year} {2010})}\BibitemShut {NoStop}%
\bibitem [{\citenamefont {Allen}, \citenamefont {Valeriani},\ and\
  \citenamefont {ten Wolde}(2009)}]{allen2009forward}%
  \BibitemOpen
  \bibfield  {author} {\bibinfo {author} {\bibfnamefont {R.~J.}\ \bibnamefont
  {Allen}}, \bibinfo {author} {\bibfnamefont {C.}~\bibnamefont {Valeriani}},\
  and\ \bibinfo {author} {\bibfnamefont {P.~R.}\ \bibnamefont {ten Wolde}},\
  }\bibfield  {title} {\enquote {\bibinfo {title} {Forward flux sampling for
  rare event simulations},}\ }\href
  {https://doi.org/10.1088/0953-8984/21/46/463102} {\bibfield  {journal}
  {\bibinfo  {journal} {J. Phys.: Condens. Matter}\ }\textbf {\bibinfo {volume}
  {21}},\ \bibinfo {pages} {463102} (\bibinfo {year} {2009})}\BibitemShut
  {NoStop}%
\bibitem [{\citenamefont {Chaudhury}\ and\ \citenamefont
  {Makarov}(2010)}]{Makarov2010Harmonic}%
  \BibitemOpen
  \bibfield  {author} {\bibinfo {author} {\bibfnamefont {S.}~\bibnamefont
  {Chaudhury}}\ and\ \bibinfo {author} {\bibfnamefont {D.}~\bibnamefont
  {Makarov}},\ }\bibfield  {title} {\enquote {\bibinfo {title} {A harmonic
  transition state approximation for the duration of reactive events in complex
  molecular arrangements},}\ }\href {https://doi.org/10.1063/1.3459058}
  {\bibfield  {journal} {\bibinfo  {journal} {J. Chem. Phys.}\ }\textbf
  {\bibinfo {volume} {133}},\ \bibinfo {pages} {034118} (\bibinfo {year}
  {2010})}\BibitemShut {NoStop}%
\bibitem [{\citenamefont {Hummer}(2004)}]{Hummer2004Transitionpath}%
  \BibitemOpen
  \bibfield  {author} {\bibinfo {author} {\bibfnamefont {G.}~\bibnamefont
  {Hummer}},\ }\bibfield  {title} {\enquote {\bibinfo {title} {From transition
  paths to transition states and rate coefficients},}\ }\href
  {https://doi.org/https://doi.org/10.1063/1.1630572} {\bibfield  {journal}
  {\bibinfo  {journal} {J. Chem. Phys.}\ }\textbf {\bibinfo {volume} {120}},\
  \bibinfo {pages} {516} (\bibinfo {year} {2004})}\BibitemShut {NoStop}%
\bibitem [{\citenamefont {Saito}(1996)}]{saito1996statistical}%
  \BibitemOpen
  \bibfield  {author} {\bibinfo {author} {\bibfnamefont {Y.}~\bibnamefont
  {Saito}},\ }\href@noop {} {\emph {\bibinfo {title} {Statistical physics of
  crystal growth}}}\ (\bibinfo  {publisher} {World Scientific, Singapore},\
  \bibinfo {year} {1996})\BibitemShut {NoStop}%
\bibitem [{\citenamefont {Jawerth}\ \emph {et~al.}(2020)\citenamefont
  {Jawerth}, \citenamefont {Fischer-Friedrich}, \citenamefont {Saha},
  \citenamefont {Wang}, \citenamefont {Franzmann}, \citenamefont {Zhang},
  \citenamefont {Sachweh}, \citenamefont {Ruer}, \citenamefont {Ijavi},
  \citenamefont {Saha}, \citenamefont {Mahamid}, \citenamefont {Hyman},\ and\
  \citenamefont {J{\"u}licher}}]{Jawerth2020aging}%
  \BibitemOpen
  \bibfield  {author} {\bibinfo {author} {\bibfnamefont {L.}~\bibnamefont
  {Jawerth}}, \bibinfo {author} {\bibfnamefont {E.}~\bibnamefont
  {Fischer-Friedrich}}, \bibinfo {author} {\bibfnamefont {S.}~\bibnamefont
  {Saha}}, \bibinfo {author} {\bibfnamefont {J.}~\bibnamefont {Wang}}, \bibinfo
  {author} {\bibfnamefont {T.}~\bibnamefont {Franzmann}}, \bibinfo {author}
  {\bibfnamefont {X.}~\bibnamefont {Zhang}}, \bibinfo {author} {\bibfnamefont
  {J.}~\bibnamefont {Sachweh}}, \bibinfo {author} {\bibfnamefont
  {M.}~\bibnamefont {Ruer}}, \bibinfo {author} {\bibfnamefont {M.}~\bibnamefont
  {Ijavi}}, \bibinfo {author} {\bibfnamefont {S.}~\bibnamefont {Saha}},
  \bibinfo {author} {\bibfnamefont {J.}~\bibnamefont {Mahamid}}, \bibinfo
  {author} {\bibfnamefont {A.~A.}\ \bibnamefont {Hyman}},\ and\ \bibinfo
  {author} {\bibfnamefont {F.}~\bibnamefont {J{\"u}licher}},\ }\bibfield
  {title} {\enquote {\bibinfo {title} {Protein condensates as aging {M}axwell
  fluids},}\ }\href {https://doi.org/DOI: 10.1126/science.aaw4951} {\bibfield
  {journal} {\bibinfo  {journal} {Science}\ }\textbf {\bibinfo {volume}
  {370}},\ \bibinfo {pages} {1317} (\bibinfo {year} {2020})}\BibitemShut
  {NoStop}%
\bibitem [{\citenamefont {Caragine}, \citenamefont {Haley},\ and\ \citenamefont
  {Zidovska}(2018)}]{Caragine2018Surface}%
  \BibitemOpen
  \bibfield  {author} {\bibinfo {author} {\bibfnamefont {C.~M.}\ \bibnamefont
  {Caragine}}, \bibinfo {author} {\bibfnamefont {S.~C.}\ \bibnamefont
  {Haley}},\ and\ \bibinfo {author} {\bibfnamefont {A.}~\bibnamefont
  {Zidovska}},\ }\bibfield  {title} {\enquote {\bibinfo {title} {Surface
  fuctuations and coalescence of nucleolar droplets in the human cell
  nucleus},}\ }\href {https://doi.org/10.1103/PhysRevLett.121.148101}
  {\bibfield  {journal} {\bibinfo  {journal} {Phys. Rev. Lett.}\ }\textbf
  {\bibinfo {volume} {121}},\ \bibinfo {pages} {148101} (\bibinfo {year}
  {2018})}\BibitemShut {NoStop}%
\bibitem [{\citenamefont {Saleh}, \citenamefont {Jeon},\ and\ \citenamefont
  {Liedl}(2020)}]{saleh2020enzymatic}%
  \BibitemOpen
  \bibfield  {author} {\bibinfo {author} {\bibfnamefont {O.~A.}\ \bibnamefont
  {Saleh}}, \bibinfo {author} {\bibfnamefont {B.-j.}\ \bibnamefont {Jeon}},\
  and\ \bibinfo {author} {\bibfnamefont {T.}~\bibnamefont {Liedl}},\ }\bibfield
   {title} {\enquote {\bibinfo {title} {Enzymatic degradation of liquid
  droplets of {DNA} is modulated near the phase boundary},}\ }\href
  {https://doi.org/10.1073/pnas.2001654117} {\bibfield  {journal} {\bibinfo
  {journal} {Proc. Natl. Acad. Sci. U.S.A.}\ }\textbf {\bibinfo {volume}
  {117}},\ \bibinfo {pages} {16160} (\bibinfo {year} {2020})}\BibitemShut
  {NoStop}%
\bibitem [{\citenamefont {Sp{\"a}th}\ \emph {et~al.}(2021)\citenamefont
  {Sp{\"a}th}, \citenamefont {Donau}, \citenamefont {Bergmann}, \citenamefont
  {Kr{\"a}nzlein}, \citenamefont {Synatschke}, \citenamefont {Rieger},\ and\
  \citenamefont {Boekhoven}}]{spaeth2021molecular}%
  \BibitemOpen
  \bibfield  {author} {\bibinfo {author} {\bibfnamefont {F.}~\bibnamefont
  {Sp{\"a}th}}, \bibinfo {author} {\bibfnamefont {C.}~\bibnamefont {Donau}},
  \bibinfo {author} {\bibfnamefont {A.~M.}\ \bibnamefont {Bergmann}}, \bibinfo
  {author} {\bibfnamefont {M.}~\bibnamefont {Kr{\"a}nzlein}}, \bibinfo {author}
  {\bibfnamefont {C.~V.}\ \bibnamefont {Synatschke}}, \bibinfo {author}
  {\bibfnamefont {B.}~\bibnamefont {Rieger}},\ and\ \bibinfo {author}
  {\bibfnamefont {J.}~\bibnamefont {Boekhoven}},\ }\bibfield  {title} {\enquote
  {\bibinfo {title} {Molecular design of chemically fueled
  peptide--polyelectrolyte coacervate-based assemblies},}\ }\href
  {https://doi.org/10.1021/jacs.1c01148} {\bibfield  {journal} {\bibinfo
  {journal} {J. Am. Chem. Soc.}\ }\textbf {\bibinfo {volume} {143}},\ \bibinfo
  {pages} {4782} (\bibinfo {year} {2021})}\BibitemShut {NoStop}%
\bibitem [{\citenamefont {Nakashima}\ \emph {et~al.}(2021)\citenamefont
  {Nakashima}, \citenamefont {van Haren}, \citenamefont {Andr{\'e}},
  \citenamefont {Robu},\ and\ \citenamefont {Spruijt}}]{nakashima2021active}%
  \BibitemOpen
  \bibfield  {author} {\bibinfo {author} {\bibfnamefont {K.~K.}\ \bibnamefont
  {Nakashima}}, \bibinfo {author} {\bibfnamefont {M.~H.}\ \bibnamefont {van
  Haren}}, \bibinfo {author} {\bibfnamefont {A.~A.~M.}\ \bibnamefont
  {Andr{\'e}}}, \bibinfo {author} {\bibfnamefont {I.}~\bibnamefont {Robu}},\
  and\ \bibinfo {author} {\bibfnamefont {E.}~\bibnamefont {Spruijt}},\
  }\bibfield  {title} {\enquote {\bibinfo {title} {Active coacervate droplets
  are protocells that grow and resist {O}stwald ripening},}\ }\href
  {https://doi.org/10.1038/s41467-021-24111-x} {\bibfield  {journal} {\bibinfo
  {journal} {Nat. Comm.}\ }\textbf {\bibinfo {volume} {12}},\ \bibinfo {pages}
  {3819} (\bibinfo {year} {2021})}\BibitemShut {NoStop}%
\bibitem [{\citenamefont {Shneidman}, \citenamefont {Jackson},\ and\
  \citenamefont {Beatty}(1999)}]{shneidman1999analytical}%
  \BibitemOpen
  \bibfield  {author} {\bibinfo {author} {\bibfnamefont {V.~A.}\ \bibnamefont
  {Shneidman}}, \bibinfo {author} {\bibfnamefont {K.~A.}\ \bibnamefont
  {Jackson}},\ and\ \bibinfo {author} {\bibfnamefont {K.~M.}\ \bibnamefont
  {Beatty}},\ }\bibfield  {title} {\enquote {\bibinfo {title} {On the
  applicability of the classical nucleation theory in an {I}sing system},}\
  }\href {https://doi.org/10.1063/1.479985} {\bibfield  {journal} {\bibinfo
  {journal} {J. Chem. Phys.}\ }\textbf {\bibinfo {volume} {111}},\ \bibinfo
  {pages} {6932} (\bibinfo {year} {1999})}\BibitemShut {NoStop}%
\end{thebibliography}
\end{document}